\documentclass[manuscript,screen]{acmart}

\renewcommand\footnotetextcopyrightpermission[1]{} 
\AtBeginDocument{%
  \providecommand\BibTeX{{%
    \normalfont B\kern-0.5em{\scshape i\kern-0.25em b}\kern-0.8em\TeX}}}

\copyrightyear{2023}
\acmYear{2023}
\acmDOI{}

\acmConference[xxx '25]{Make sure to enter the correct conference title from your rights confirmation email}{xxx}{xx}
\acmPrice{}
\acmISBN{}
\usepackage{amsmath}
\usepackage{graphicx} 		%图片
\usepackage{subfigure}		%子图
\usepackage{makecell}
\usepackage{colortbl}
\usepackage{array}
\usepackage{multirow}

\usepackage{caption}
\begin{document}
%%
%% The "title" command has an optional parameter,
%% allowing the author to define a "short title" to be used in page headers.

\title[DeMod]{DeMod: A Holistic Tool with Explainable Detection and Personalized Modification for Toxicity Censorship}

%%
%% The "author" command and its associated commands are used to define
%% the authors and their affiliations.
%% Of note is the shared affiliation of the first two authors, and the
%% "authornote" and "authornotemark" commands
%% used to denote shared contribution to the research.

%\authornote{Both authors contributed equally to this research.}

\author{Yaqiong Li}
\email{liyq22@m.fudan.edu.cn}
\affiliation{%
  \institution{Fudan University}
  \city{Shanghai}
  \country{China}
}

\author{Peng Zhang}
%\authornotemark[1]
\email{zhangpeng_@fudan.edu.cn}
\affiliation{%
  \institution{Fudan University}
  \city{Shanghai}
  \country{China}
}

\author{Hansu Gu}
\email{guhansu@gmail.com}
\affiliation{%
    \institution{Independent}
  \city{Seattle}
  \country{USA}
}

\author{Tun Lu}
\email{lutun@fudan.edu.cn}
\affiliation{%
  \institution{Shanghai Key Laboratory of Data Science}
   \city{Shanghai}
   \country{China}
}
\affiliation{%
  \institution{Fudan University}
  \city{Shanghai}
  \country{China}
}

\author{Siyuan Qiao}
\email{syqiao23@m.fudan.edu.cn}
\affiliation{%
  \institution{Fudan University}
  \city{Shanghai}
  \country{China}
}

\author{Yubo Shu}
\email{ybshu20@fudan.edu.cn}
\affiliation{%
  \institution{Fudan University}
  \city{Shanghai}
  \country{China}
}

\author{Yiyang Shao}
\email{yyshao22@m.fudan.edu.cn}
\affiliation{%
  \institution{Fudan University}
  \city{Shanghai}
  \country{China}
}

\author{Ning Gu}
\email{ninggu@fudan.edu.cn}
\affiliation{%
  \institution{Fudan University}
  \city{Shanghai}
  \country{China}
}

%%
%% By default, the full list of authors will be used in the page
%% headers. Often, this list is too long, and will overlap
%% other information printed in the page headers. This command allows
%% the author to define a more concise list
%% of authors' names for this purpose.
\renewcommand{\shortauthors}{Yaqiong Li et al.}

%%
%% The abstract is a short summary of the work to be presented in the
%% article.
\begin{abstract}
Although there have been automated approaches and tools supporting toxicity censorship for social posts, most of them focus on detection. Toxicity censorship is a complex process, wherein detection is just an initial task and a user can have further needs such as rationale understanding and content modification. For this problem, we conduct a needfinding study to investigate people's diverse needs in toxicity censorship and then build a ChatGPT-based censorship tool named DeMod accordingly. DeMod is equipped with the features of explainable \textbf{De}tection and personalized \textbf{Mod}ification, providing fine-grained detection results, detailed explanations, and personalized modification suggestions. We also implemented the tool and recruited 35 Weibo users for evaluation. The results suggest DeMod's multiple strengths like the richness of functionality, the accuracy of censorship, and ease of use. Based on the findings, we further propose several insights into the design of content censorship systems.

\end{abstract}
%%
%% The code below is generated by the tool at http://dl.acm.org/ccs.cfm.
%% Please copy and paste the code instead of the example below. 
%%
\ccsdesc[500]{Human-centered computing~Collaborative and social computing}
%%
%% Keywords. The author(s) should pick words that accurately describe
%% the work being presented. Separate the keywords with commas.
\keywords{toxicity censorship, explainable detection, personalized modification, ChatGPT}

\maketitle

\section{Introduction}
Nowadays, social media sites have been popular mediums for self-disclosure. For example, hundreds of millions of people utilize Twitter \cite{Jhaver_60}, Facebook \cite{Seering_3, Pan_19, Saha_37}, and Weibo \cite{Zhao_86} to record life events, express personal thoughts and opinions, and interact with friends every day. The openness of social media provides a spacious environment for content sharing while resulting in the disclosure of toxic content (toxicity), defined as "\emph{a rude, disrespectful, or unreasonable comment that is likely to make someone leave a discussion}" \cite{Perspective_140}, including hate speech \cite{Kang_50}, harassment \cite{Jhaver_60, Cai_68}, insults and abuse \cite{Beres_34}, and offensive language \cite{deng2022cold_138}, etc. Since the severe problem of context collapse \cite{Misra_152}, social media users are usually unaware of the disclosure of toxic content. For example, the prior studies \cite{wiegand2021implicitly_156, sap2019social_160} found that about two-thirds of toxic content was implicit toxicity in online communities and the corresponding users were usually unaware of the content and the harm to others. Research revealed that 23.00\% of users regret when they re-examine their shared content due to several reasons \cite{wang2011regretted_155}, such as lack of the consequence consideration of posts, culture misjudgment, unintended audience, misunderstanding of platform norms.

To avoid toxic content disclosure, social media users generally conduct content censorship before publishing a post. The censorship procedure can be implemented by users themselves or by leveraging some automated tools. For example, several studies have found that individuals usually censored their content by checking, adjusting, or even deleting part of the content to make the content suitable to be published on social media \cite{wisniewski2012fighting_158}. Although there have been various censorship approaches, most of them focus on toxic content detection, e.g., toxicity score evaluation with Perspective API \cite{Perspective_140} and toxic keywords identification \cite{Wright_48}.
Toxic content censorship is a complex process, wherein detection is just an initial task, and a user can have diverse needs such as detection result understanding and content modification. For example, a user can identify toxic words in the content with the RECAST tool \cite{Wright_48} while not knowing how to reduce its toxicity limited by her/his knowledge or experience. Therefore, there needs a holistic automated tool that can help social media users conduct multiple censorship tasks including toxic content detection, content modification, etc. 

Building a holistic tool for toxicity censorship faces several challenges. First, social media users' diverse needs for toxic content censorship remain unknown. As mentioned, social media users may have different function demands like enriching explanations and giving modifications. Therefore, a systematic investigation of toxic content censorship demands is needed when conducting research on a holistic censorship tool, aggravating the complexity of this study. Second, designing and implementing a toxicity censorship tool that meets the diverse needs of users is non-trivial. Such a tool should be characterized by multiple objectives like accurate detection, fine-grained results, and appropriate revisions. How to achieve different functions and integrate them efficiently is a challenging task. Third, extensive evaluations in practice are difficult to conduct. To demonstrate the tool's performance in helping users censor toxic content, it needs to conduct long-term evaluations in real social media scenarios by using various measurements, while some of them like the modification effects, are difficult to measure.

For the above problem and challenges, we explore to design a holistic automated tool for helping users conduct toxic content censorship on social media. 
First, we conduct a needfinding study on a popular Chinese social platform - Weibo to systematically understand users' current toxicity censorship practice, the problems encountered, and their corresponding expectations for system design. By combining a questionnaire survey and interviews, we uncover users' diverse demands for the design of toxicity censorship tools and propose five goals to guide our system design, including providing holistic censorship, offering fine-grained detection results, strengthening interpretability, giving personalized revising suggestions, and ensuring user-control. 
Second, according to these goals, we design and implement a holistic automated toxicity censorship tool named DeMod. It is essentially a ChatGPT-enhanced tool equipped with the modules of explainable \textbf{De}tector and personalized \textbf{Mod}ifier. The explainable Detector can detect toxic content by giving fine-grained results like keywords and providing immediate and dynamic explanations. The immediate explanation clarifies why the content is toxic, and the dynamic explanation simulates audiences' attitudes to the forthcoming post, helping a user know the content's potential effects. Both explanations aim to enhance the user's understanding of toxic content and encourage behavior regulation. After that, the modifier gives suggestions on how to revise the toxic content by considering multiple requirements, including detoxifying, reserving the original semantics, and revealing a user's personalized language style. By taking advantage of these modules, social media users can conduct content censorship more efficiently and flexibly. 
Third, we implement DeMod as a third-party tool by setting Weibo as a research site and recruit 35 participants to conduct extensive evaluations. We adopt several metrics regarding our design goals. The evaluation results suggest DeMod's capability in toxicity censorship and high acceptance among participants. Based on the above work and results, we also propose several insights into the design of content censorship tools, including enhancing censorship tools from the holistic perspective, emphasizing the interpretability of the process and results, and providing improvement measures to assist users in posting better. To conclude, our contributions can be summarized as:

\begin{itemize}
    \item We conduct a needfinding study to investigate social media users' current toxicity censorship practice, the problems encountered, and their corresponding expectations for system design, based on which five design goals are proposed to guide the improvement of toxicity censorship tools.    
    \item  We propose a holistic automated tool based on ChatGPT for helping users conduct toxicity censorship. To the best of our knowledge, this is the first work that supports users' demands in multiple stages of toxicity censorship beyond detection. 
    \item We conduct extensive evaluations in real social media scenarios and validate DeMod's strengths in toxicity censorship.   
    \item Several insights are proposed for the further improvement of content censorship system design.
\end{itemize}
The rest of this paper is organized as follows. In Section 2, we review related research on content censorship and large language models. In Section 3, we introduce the procedure and results of our empirical study. The framework of DeMod and its implementation are given in Section 4. Section 5 exhibits our evaluation settings and results. Section 6 discusses our findings, and the limitations and future work are clarified in Section 7. Finally, conclusions are given in Section 8.

\section{Related Work}
\subsection{Content Censorship}
In the social media context, users generally conduct content censorship (also called “last-minute self-censorship” \cite{das2013self_153}) by themselves or employing automated tools. A study indicated that individuals generally manually censored their content before sharing by checking, adjusting, or deleting some words to ensure the content's consistency with platform norms and cultures \cite{Aghajari_26, cho2020will_154}. However, this self-censorship process relies heavily on users' knowledge, experience, and time, affecting censorship efficiency and quality. So, studies have emerged focusing on building automated tools that help users facilitate content censorship. For example, users can use some third-party tools \cite{Perspective_140} to identify toxic content from their posts, including hate speech \cite{Kang_50}, harassment \cite{Jhaver_60, Cai_68, Blackwell_72, Nova_78}, insults and abuse \cite{Beres_34}, etc. 
Although there have been diverse censorship tools, most of them focus on detection without considering users' composite censorship demands like result understanding and content modification. These problems result in the low efficiency of users' content censorship. So many users choose to publish content without censorship but rely on platforms' moderation measures.

Different from censorship, content moderation is initialized by a social media platform \cite{content_163}, with the aim of monitoring whether content submitted to the platform complies with the platform's rules and guidelines \cite{Aghajari_26, cho2020will_154}. So content moderation generally occurs right after content publishing and is also called post-moderation \cite{content_162}. Although content moderation can also identify toxic content, social media users are generally passive in this procedure and toxic content has resulted in some impacts when being moderated \cite{Vaccaro_1}.
Many previous studies have explored automated \cite{Lai_31, Song_49, Kiene_47} and human moderation approaches \cite{Cai_5, McInnis_13, Kou_35, Schluger_17} and their moderation targets are similar to that of content censorship. For example, \cite{Stratta_44} designed a Chrome extension program that automatically generates content warnings by utilizing keyword identification and online intervention interface principles, aiming to identify sensitive information in contents. \cite{Jhaver_32} presented a word filter to detect some harmful comments, like harassment \cite{Jhaver_60, Cai_68, Blackwell_72, Nova_78} and targeted abuse \cite{Beres_34}. Moreover, \cite{Wright_48} developed an open-sourced visualization tool to identify toxic content and predict the toxicity score of keywords, helping human moderators improve moderation efficiency and accuracy. 

Compared to platform post-moderation, content censorship is a user-driven and pre-check process. It has several benefits, including instant feedback, autonomous control over contents \cite{wei2023there_166}, and proactive checking avoiding potential social impact \cite{jamil2018green_168}. Like “we [the HCI \& security communities] have used user effort as a first resort, not last” \cite{herley2013more_167}, the instant feedback of censorship by users allows them to promptly gauge potential issues with their content, providing an opportunity for adjustments or edits. Although the personalized censorship on post's privacy publicity is studied \cite{Liu_84}, there still lacks the relevant work of user autonomy over their personal expression under platform moderation criteria \cite{wei2023there_166}. Besides, content censorship enables users to avoid posting material that could negatively impact their personal reputation \cite{jamil2018green_168}, and it also minimizes the likelihood of subsequent platform moderation or punitive measures, promoting a smoother online experience. 

\subsection{Toxicity Detection with/for LLMs}
There have been extensive works on exploring methods for detecting toxic content \cite{Perspective_140, Wright_48}, including hate speech \cite{Kang_50}, harassment \cite{Jhaver_60, Cai_68}, insults and abuse \cite{Beres_34}, offensive language \cite{deng2022cold_138}, etc. These methods are proposed with various models and algorithms, such as feature-based classifiers \cite{ibrohim2019multi_171}, neural network architectures (CNN, LSTM, etc.) \cite{sigurbergsson_174}, and pre-trained language models (BERT \cite{devlin2018bert_94}, RoBERTa \cite{liu2019roberta_139}, etc.). Previous studies have also explored other strategies for accuracy improvement in toxicity detection, including constructing high-quality corpus \cite{deng2022cold_138}, designing detailed classification principles \cite{elsherief-etal-2021_176}, etc. For example, a model named HateBERT is re-trained based on BERT for abusive language detection \cite{hatebert_175}. It utilizes a training corpus derived from Reddit, which involves offensive, abusive, and hateful content. Besides the above research and practice, a comprehensive framework has been proposed for toxicity detection \cite{SHETH2022312_184}, which incorporates contextual knowledge such as semantics, intent, and sentiment. However, the framework remains theoretical and necessitates further deliberation and analysis in both information collection and validation.

Recently, the emergence of Large Language Models (LLMs) has brought impressive effects in promoting NLP research and applications, especially ChatGPT \cite{ChatGPT_93}. There are some efforts in toxicity detection regarding LLMs \cite{zhang2023efficient_170,mishra2023exploring_169,nguyen2023fine_149}, including toxicity detection for LLMs and toxicity detection employing LLMs. The former focuses on LLMs' outputs, aiming to avoid the appearance of toxic content in LLMs' generations. The latter employs LLMs as a tool to detect toxic content by taking advantage of LLMs' strong comprehension and reasoning abilities, which can avoid the tedious procedure of feature engineering and model training. For example, Llama2 \cite{touvron2023llama_95} has been used to detect online sexual predatory chats and abusive texts with fine-tuning techniques \cite{nguyen2023fine_149}, wherein traditional processing such as feature extraction (semantics, intent, or sentiment) is no longer needed. Another work explored ChatGPT's performance in detecting toxic comments on GitHub and designed various prompts to justify model outputs \cite{mishra2023exploring_169}. A novel prompt design approach \cite{zhang2023efficient_170}, named Decision-Tree-of-Thought, was also proposed to guide LLMs in enhancing the quality of toxicity detection. 

\subsection{Our Work in Context}
Toxicity detection is just an initial task in toxic content censorship and users can have other diverse demands in the process. Therefore, our work aims to build a holistic automated toxicity censorship tool with the benefit of LLMs, addressing social media users' diverse problems and expectations in posting. We first investigate the problems users encountered in toxicity censorship and their expectations for handling them. Based on that, we propose a novel multi-functional censorship tool based on ChatGPT. The tool is equipped with multiple features like toxicity detection, result explanation, and content modification. To the best of our knowledge, this is the first tool that supports users' demands in multiple stages of toxic content censorship.

\section{NeedFinding Study} 
We began with a needfinding study to understand the current toxicity censorship practices of social media users, including how to conduct toxic content censorship, the problems encountered, the corresponding expectations, etc.
\subsection{Method}
Our needfinding study was conducted on a popular Chinese social media site - Weibo \cite{han2023public_161}, and both questionnaires and semi-structured interviews were adopted. We chose Weibo as our research platform for several reasons. First, Weibo has a large user base with 600 million monthly active users \cite{Weibofinancialreport_191}, and its post content involves diverse topics, including personal life, hot events, entertainment, etc. Second, a large amount of toxic content is generated and disseminated on Weibo, and the categories of toxicity are diverse \cite{weibo_zhili_163}, including harassment, offensive language, insult and abuse, etc., which are similar to those of other popular social media platforms like Twitter and Instagram. For example, from November 2022 to August 2023, the number of offensive expressions identified on Weibo exceeded 120 million \cite{weibo_zhili_163}. Thus, Weibo has been a common-used Chinese research platform for toxicity studies \cite{deng2022cold_138, Xiang2020CironAN_187, Jiang2022SexWEsDW_188}. 

For our needfinding study, we initially used questionnaires to investigate the current practices of toxicity censorship among Weibo users and identify the problems they encountered. Subsequently, semi-structured interviews were employed to find users' desired design features for toxicity censorship. The findings can guide us to design a human-centered tool to improve users' toxicity censorship practices.

\textbf{Participants.} We released the questionnaire on an online survey platform and posted it to social media. A total of 493 participants (214 males and 279 females, aged between 15 and 58) finished the questionnaire. Most (439, 89.05\%) of these participants are aged between 18 and 35 years old, and people of this age range are the primary users of social media. Among these participants, 30 persons (18 males and 12 females, aged between 18 and 35) expressed their willingness to participate in the subsequent semi-structured interviews.

\textbf{Procedure.} The aim of the questionnaire is to investigate the current practices of toxic content censorship among Weibo users and the problems they encountered. It comprises three parts with a total of 11 questions (5-point Likert scale): 1) a user's basic profiles, such as the demographic features (gender, age, etc.), the habit of Weibo use, and post frequency; 2) current censorship practices, including whether usually conducting toxicity censorship on Weibo and how to censor; 3) problems encountered during content censorship.
To further understand users' expectations of toxic content censorship, we designed a draft framework and invited 30 participants to conduct participatory design through semi-structured interviews with offline or online meetings. Participants expressed their expectations for the tool's design and outputs, including the detection granularity (binary or multi-classification), object granularity (post, sentence, or word), etc. According to participants' feedback, we iteratively adjusted our design. With the participants' agreement, the interviews were recorded and then transcribed by automatic tools and the first author. All of the data would not be shared to avoid privacy leakage. 

Referring to common procedure \cite{TangiblePrivacyUserCentric2020_178, ReliabilityInterraterReliability2019_179}, we took an analysis of participants' feedback and interview logs, using statistical analysis and Thematic Analysis methods \cite{marchWikipediaEditathonsSites2020_181}. Initially, three authors performed open coding on all participants' feedback independently and then worked together to build a series of axial codes. Following this, the authors reviewed the interview logs, iteratively refining the coding scheme across three rounds to address shortcomings in the previous round. The final stage involved focused coding, aimed at synthesizing evolving conceptual categories into more comprehensive topics related to censorship experiences like detection or display requirements. Throughout the whole process, three authors kept communication regularly with other authors, ensuring conceptual coherence and reliability. The coding process was deemed complete upon reaching a consensus among the authors on the conclusions.

\subsection{Results} 
\textbf{Most users tend to prevent posting toxicity through two approaches: self-censorship and platform moderation.} According to the results of our questionnaire, 355 participants (71.60\%) use Weibo. 227 participants (63.94\%) publish posts on Weibo, wherein 11.27\% publish daily, 20.85\% weekly, and 30.7\% monthly, indicating their high activity levels on the platform. Most users choose to conduct toxicity censorship when posting, and only few (21, 9.25\%) mentioned never. This phenomenon demonstrates people's strong awareness of content censorship on Weibo. Among the 156 participants who provided the answer of censorship ways generally used, 112 participants (72.44\%) selected self-censorship (censoring posts by oneself), 91 (58.33\%) selected relying on Weibo's platform moderation (identifying and removing the posts that violate platform norms \cite{Cai_5, Scheuerman_6}), and 30 (19.23\%) selected inviting others to censor (seeking advice from parents, friends, or other individuals). 

\textbf{Problems of current censorship approaches.} Among the 112 participants choosing self-censorship, only 11 (8.53\%) thought it could meet their needs, and the problems can be summarized as the lack of censorship accuracy and objectivity due to the users' limited knowledge, experience, and time. 15 participants acknowledged this phenomenon, saying "\emph{It is always influenced by my subjective understanding}", "\emph{I don't know if something is nontoxic sometimes}", "\emph{Maybe my knowledge is not enough to determine which word violates the platform norms}", etc. These responses suggest self-censorship heavily relies on users' knowledge, experience, and time, and users themselves cannot perform accurately and objectively. For the platform moderation on Weibo \cite{Zhao_86}, only 4 participants (4.60\%) believed this approach could meet their censorship needs, and the main problems can be summarized as the lack of user control (moderation occurs after posts have been posted and users cannot take some proactive actions), lack of explanation (why posts are toxic), and low accuracy. 

\textbf{Design Goals.} We further analyzed participants' expectations for the design of toxicity censorship tools, based on which the following five design goals are proposed. 

\begin{itemize}
    \item \textbf{G1: Provide holistic censorship.} Toxicity censorship tools should provide holistic functions, including toxicity detection and modification. All participants confirmed the necessity of such a tool, saying "\emph{This tool can greatly unload my brain. I am often not aware my words may hurt others}", "\emph{I just post and wish a holistic tool that can point out my issues and offer revision suggestions}", etc. It can not only alleviate users' censorship burden but also improve the accuracy and objectivity.
    
    \item \textbf{G2: Offer fine-grained detection results.} Toxicity censorship tools should provide fine-grained detection results. Not only a classification result (whether a post contains toxicity) but also fine-grained results (the related sentences, phrases, and words) should be given in toxicity censorship to promote user perception. Participants mentioned, "\emph{It's not enough to tell me whether my post is toxic or not. The specific words or phrases that might harm others should be identified}", "\emph{The keywords should be highlighted directly. I don't want to waste my time, it's just a post}", etc. 
    
    \item \textbf{G3: Strengthen interpretability.} Participants wish for an immediate explanation about detection results, saying "\emph{For the toxic content that I may not realize, it's better to offer some reasons to let me know whether I should post the content}", "\emph{Highlighted words would be clear and intuitive}", etc. Moreover, 26 participants (86.87\%) expressed a desire to get to know audiences' views on the posts proactively, helping them understand why some posts cannot be published. For example, P5 responded, "\emph{There are usually people who are not satisfied with my words, and I don't like to be preached by others either. This feature [getting to know the potential social impacts of the post content] is quite interesting, as it allows me to know whether my words or expression has any issues during conversations}".
    
    \item \textbf{G4: Give personalized revising suggestions.} To alleviate users' modification burden, the toxic content censorship tool should give revising suggestions that can make the post content normal while remaining the original semantics and a user's personalized language style in the meanwhile. 28 participants (93.33\%) expressed the thought to reduce their modification burden. During modification, semantics and personalized language style should be reserved as much as possible after revision. For example, participants said, "\emph{If there are some inappropriate sentences or words, it would be better to replace with some subtly expressions automatically and I don't want to edit directly}" and "\emph{I value my usual speaking style. If the revision is too formal, there is no need to appear on my social media}".
    
    \item \textbf{G5: Ensure user-control.} In order to ensure users do not feel overly censored, content censorship should be conducted with user-awareness and user-control. What role a censorship tool plays is to give suggestions and actions like whether to revise depending on users' decisions. Participants expressed, "\emph{The function of automatic modification should give some suggestions, not publishing directly. I prefer to revise on my own}", "\emph{I prefer to use different functions at any time. Sometimes, detection is enough}", etc.
\end{itemize}

\section{DeMod: A Holistic Tool for Toxicity Censorship}
According to the design goals outlined in the previous section, we designed a tool named DeMod to help social media users conduct toxic content censorship proactively. Based on ChatGPT \cite{ChatGPT_93}, DeMod is designed to provide multiple functionalities, including explainable \textbf{de}tection and personalized \textbf{mod}ification. To demonstrate how to deploy DeMod in practice, we also implemented DeMod as a third-party tool by setting a famous social media site - Weibo as a research site. The following gives the details of DeMod's design and implementation.

\subsection{DeMod}
\begin{figure}[h]
  \centering
  \includegraphics[width=280pt]{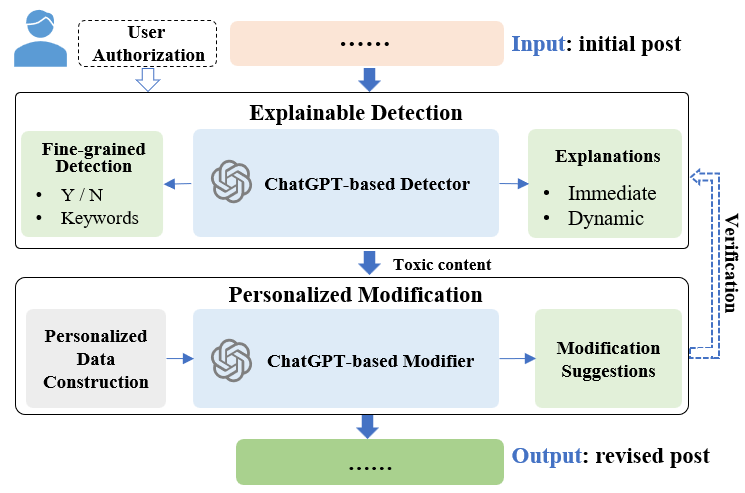}
  \caption{The framework of DeMod.}
  \label{fig:framework}
\end{figure}

According to our design goals, DeMod, presented in Figure \ref{fig:framework}, is built with three main modules.
\begin{itemize}
    \item User Authorization: This module is utilized to get a user's permission for Weibo profile access, such as the user's historical posts and social connections.
    \item Explainable Detection: This module conducts toxicity detection and explaining based on ChatGPT. Firstly, it provides multi-granularity detection results, including classification (Y/N representing if a post is or isn't toxic content, respectively) and corresponding keywords. Secondly, it gives detailed explanations, including immediate and dynamic explanations for the user. The former directly explains why certain keywords are toxic, and the latter predicts the audiences' attitudes to the post to help the user perceive the potential effects.
    \item Personalized Modification: This module provides the user with revising suggestions to avoid toxicity posting while attempting to retain the original semantics and the user's personalized language style in the meanwhile.
\end{itemize}

\subsubsection{User Authorization} DeMod is a third-party tool helping people conduct content censorship on social media. Since there are diverse personal profiles in the social media context, we first introduce the user authorization module into DeMod to avoid privacy leakage. It can be implemented by calling the OAuth API supplied by social media platforms. During authorization, a user can see what kinds of profiles (public historical posts and social connections) DeMod will use and give the consent. These profiles enable DeMod to provide more precise and personalized censorship results. The user can revoke authorization to avoid personal information misuse. 

\subsubsection{Explainable Detection} Our empirical study indicates that users wish for fine-grained detection results and interpretability in toxicity censorship. Therefore, we design DeMod with a ChatGPT-based toxicity detector to provide multi-granularity detection results and corresponding explanations. It firstly outputs the toxicity detection results, including classification (Y/N, whether a post contains toxicity) and the corresponding keywords. The classification result informs users whether a post contains toxic content. If so, keywords triggering toxicity will be highlighted to enhance users' perception of toxic details. For these detection results, the detector then gives immediate and dynamic explanations. Immediate explanation illustrates why the post and keywords are toxic. For the dynamic explanation, the detector can predict audiences' attitudes or opinions to the post by taking advantage of ChatGPT's capability in character simulation, helping users get to know the potential social impact of the content. This design aligns with the theory of basic human values, emphasizing that individual behavior is easily influenced by the values held by others \cite{hasan2023psychometric_164, schultz2005values_165}, including others' attitudes, personal information, etc. To ensure user control, DeMod allows users to manually select some audiences, like parents, friends, and even strangers, and conduct attitude simulation. Once an audience is selected, DeMod simulates her/his attitudes to the post. 

The above detection and explanation tasks are all achieved based on ChatGPT, and the prompts are shown in Appendix Figure \ref{Detector' prompts}. Both prompts contain four key elements: task description, prompt template, system setting, and output format constraints. For the first prompt in Figure \ref{Detector' prompt1}, the "Task" is "Toxicity detection", the "Prompt template" incorporates the input sentence to be detected and the relevant topic, and the "System" describes what ChatGPT should do and the corresponding requirements, including task requirements and output format. The output format is set as JSON to ensure it can be easily parsed. For the dynamic explanation task, we collect the interaction context between the current user and the selected audience (post comments the current user obtained from the selected audience on Weibo) with user consent and aggregate it as a corpus to support ChatGPT conducting attitude simulation, revealing the audience's preferences, opinions, and thoughts. Similar to the above, Figure \ref{Detector' prompt2} exhibits the prompt of the dynamic explanation task. The "Task" is "Viewpoint simulation", and the interaction context is embedded between "The start of the interaction context between the user and the selected role" and "The end of the interaction context between the user and the selected role". Moreover, the "System" setting gives the task requirements, dialogue round limits, expression style, the rules without the context (if there is no interaction between the current user and the selected audience), and output format. 

\subsubsection{Personalized Modification.} Our empirical study suggests that users wish modification suggestions to help them detoxify posts while the semantics of the original post and a user's personalized language style should be retained as much as possible. We find directly using a prompt similar to the detection prompts to let ChatGPT conduct the toxic content modification task challenging since the multiple goals cannot be achieved simultaneously. However, ChatGPT is capable of learning from a few examples, i.e., few-shot learning \cite{Brown_92}. If there are some pairs of examples exhibiting the original posts (a post with toxic content) and the corresponding revised contents (the corresponding post without toxic content but with similar semantics and the user's language style), ChatGPT can achieve these modification goals better. So, we first attempt to construct these pairs of examples as follows.

\begin{itemize}
\item {\bfseries Step 1}: For the current user, several pairs of examples $(NT_i, T_i)$ are constructed based on her/his historical posts on Weibo, wherein $NT_i$ represents a nontoxic historical post, $T_i$ represents the corresponding toxic post constructed by us, and $i$ indicates the $i$th pair ($i \in \{1,2,\cdots\}$). Both $NT_i$ and $T_i$ are characterized by the similar semantic and language style. 

\item {\bfseries Step 2}: Transform each pair into $(T_i, NT_i)$, and construct a prompt by using several pairs of these examples to stimulate ChatGPT to achieve the multiple modification goals.

\item {\bfseries Step 3}: Verify the revised post to ensure the modification's effect.
\end{itemize}

For a user, Step 1 can be executed in advance and then persisted to support Step 2 and Step 3. After that, when a user requires post-modifying in practice, only Step 2 and Step 3 are needed. The details of these three steps are described below.

\begin{figure}[h]
  \centering
  \includegraphics[width=400pt]{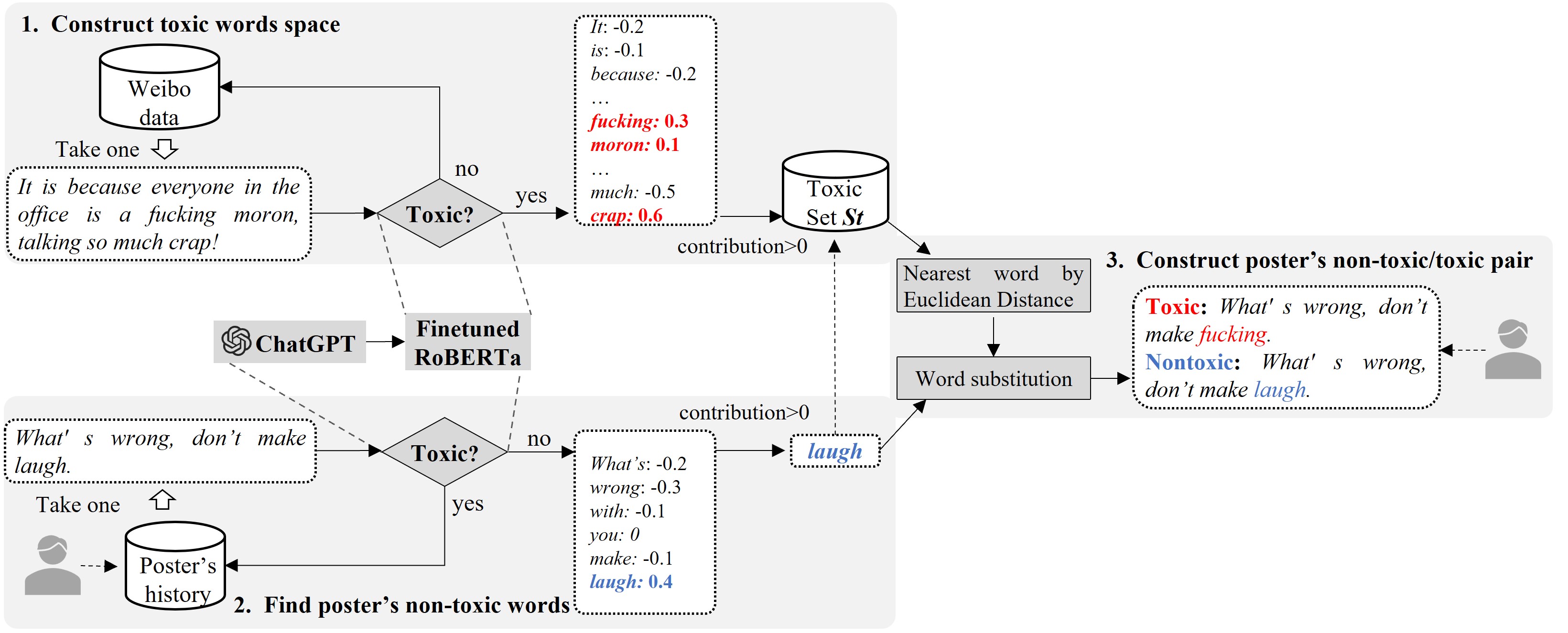}
  \caption{Personalized data construction.}
  \label{Data construction}
\end{figure}

In Step 1, we construct a post $T_i$ for each post $NT_i$ according to the word substitution strategy utilized in \cite{Wright_48}. As is shown in Figure \ref{Data construction}, the process is as follows:
\begin{enumerate}
\item {\bfseries Construct toxic word space}. Firstly, to ensure the toxic word space aligns with the features of Weibo posts, we crawled a large number of Weibo posts as a training corpus, including 4,832 users and 968,503 posts (each post is denoted as $D_i$, $i\in \{1,2,\cdots\}$). Then we used the RoBERTa model \cite{liu2019roberta_139} fine-tuned by the Chinese offensive dataset COLD \cite{deng2022cold_138} to detect toxic posts from this corpus. If a post $D_i$ is judged as toxic, calculate the contribution value $V_{ij}$ of each $token_j$ to the result, as shown in the formula \ref{eq:1} ($j$ represents the $j$th token in $D_i$). The contribution value $V_{ij}$ is then normalized to $[-1,1]$. If $V_{ij}>0$, the $j$th token $token_j$ of post $D_i$ is added to the toxic word space $S_t$. Iterate this process to traverse $D_i$.

\begin{equation}
V_{ij}=contribution(RoBERTa(D_i)=1).
\label{eq:1}
\end{equation}

\item {\bfseries Find nontoxic posts and corresponding contribution words from the current user's historical posts}. Given the current user's historical posts (each post is denoted as $H_i$, $i\in\{1,2,\cdots\}$), we utilize both ChatGPT and fine-tuned RoBERTa to identify the nontoxic posts. ChatGPT conducts toxicity detection first. If ChatGPT judges a post as nontoxic, RoBERTa will be introduced to double-check the result and further evaluate the contribution value $V_{ij}$ for each $token_j$ of $H_i$, as shown in the formula \ref{eq:2}. The reason for using both ChatGPT and RoBERTa is to ensure the selected posts are nontoxic. 

\begin{equation}
V_{ij}=contribution(RoBERTa(H_i)=0,{\rm if}\ ChatGPT(H_i)=0).
\label{eq:2}
\end{equation}

\item {\bfseries Construct nontoxic-toxic pairs for the current user}. For each of the obtained nontoxic posts, we conduct word vector mapping for each token and get the token's vector $[F_1, F_2,\cdots, F_n]$ and then find the closest vector $[F_1^{'}, F_2^{'},\cdots, F_n^{'}]$ from the toxic word space $S_t$ based on Euclidean distance, where $n$ denotes the vector dimension. The relevant loss function is presented in the following equation \ref{eq:3}. The contribution words of nontoxic posts are then substituted with the corresponding nearest toxic tokens to make nontoxic posts become toxic, based on which we can obtain several nontoxic-toxic pairs of $(NT_i, T_i)$. 

\begin{equation}
[F_1^{'},F_2^{'},\cdots,F_n^{'}]=argmin_{[F_1,F_2,\cdots,F_n]\in S_t}\sum_{i=1}^{n}(F_i-F_i^{'})^2.
\label{eq:3}
\end{equation}
\end{enumerate}

In Step 2, We flip each nontoxic-toxic pair $(NT_i, T_i)$ to $(T_i, NT_i)$ and embed them between "The start of nontoxic-toxic samples" and "The end of nontoxic-toxic samples" in the prompt shown in Appendix Figure \ref{Modifier's prompt}. The prompt has a similar format to our toxicity detection prompts. The "Task" is "Expression modification", and the "System" setting gives the task requirements, the rules without posting history (if there are no historical posts of the current user), the constraint of expression style, output format, etc. For those users without history posts, $(T_i, NT_i)$ is none. So, we add the basic examples in the "System" setting. Using this prompt to interact with ChatGPT, the modification content will be generated. 

In Step 3, a verification is conducted to confirm that toxic content has been effectively eliminated. We re-detect the modified content through the explainable detection module. If a post is toxic, DeMod will modify it again. This procedure serves to guarantee the modification's effectiveness, offering a more reliable censorship result. 

Above all, the features of DeMod can be summarized as follows regarding the design goals described in Section 3.2.

{\bfseries Holistic censorship (G1)}. 
DeMod provides not only toxicity detection but also result explanation and personalized modification for users' toxicity censorship. To the best of our knowledge, this is the first tool that can cover the multiple stages of toxicity censorship.

{\bfseries Fine-grained detection results (G2)}. When conducting toxicity detection, DeMod presents users with the classification result and corresponding keywords, enhancing users' perception of toxicity.

{\bfseries Immediate and dynamic explanations (G3)}. The immediate explanation component within the toxicity detection module provides users with clear justifications and criteria for DeMod's decision. By taking advantage of the dynamic explanation based on viewpoint simulation, users can gain insights into different audiences' attitudes. It can empower users to actively perceive the potential consequences and impacts of their posts, assisting them in making more reasonable decisions. 

{\bfseries Personalized modification (G4)}. By learning from our constructed corpus, DeMod can help users detoxify posts while preserving the original semantics and personalized language style in the meanwhile. It aligns with users' expectations for multi-objective modification.

{\bfseries User-control (G5)}. During the usage, no operations are conducted without user choices to ensure DeMod is user-driven. Even some modules like the personalized modification component can give some suggestions, the decisions will eventually be made according to user choices.

\subsection{System Implementation}
We implement DeMod as a web application through Flask and Vue, using Redis to store data on Ali cloud server~\footnote{\url{https://cn.aliyun.com}}. The following gives the tool's usage in detail.

\begin{figure*}[h]
\centering
\subfigure[Input ID or nickname]{\label{fig:input_id}
\includegraphics[width=0.31\linewidth,height=0.22\linewidth]{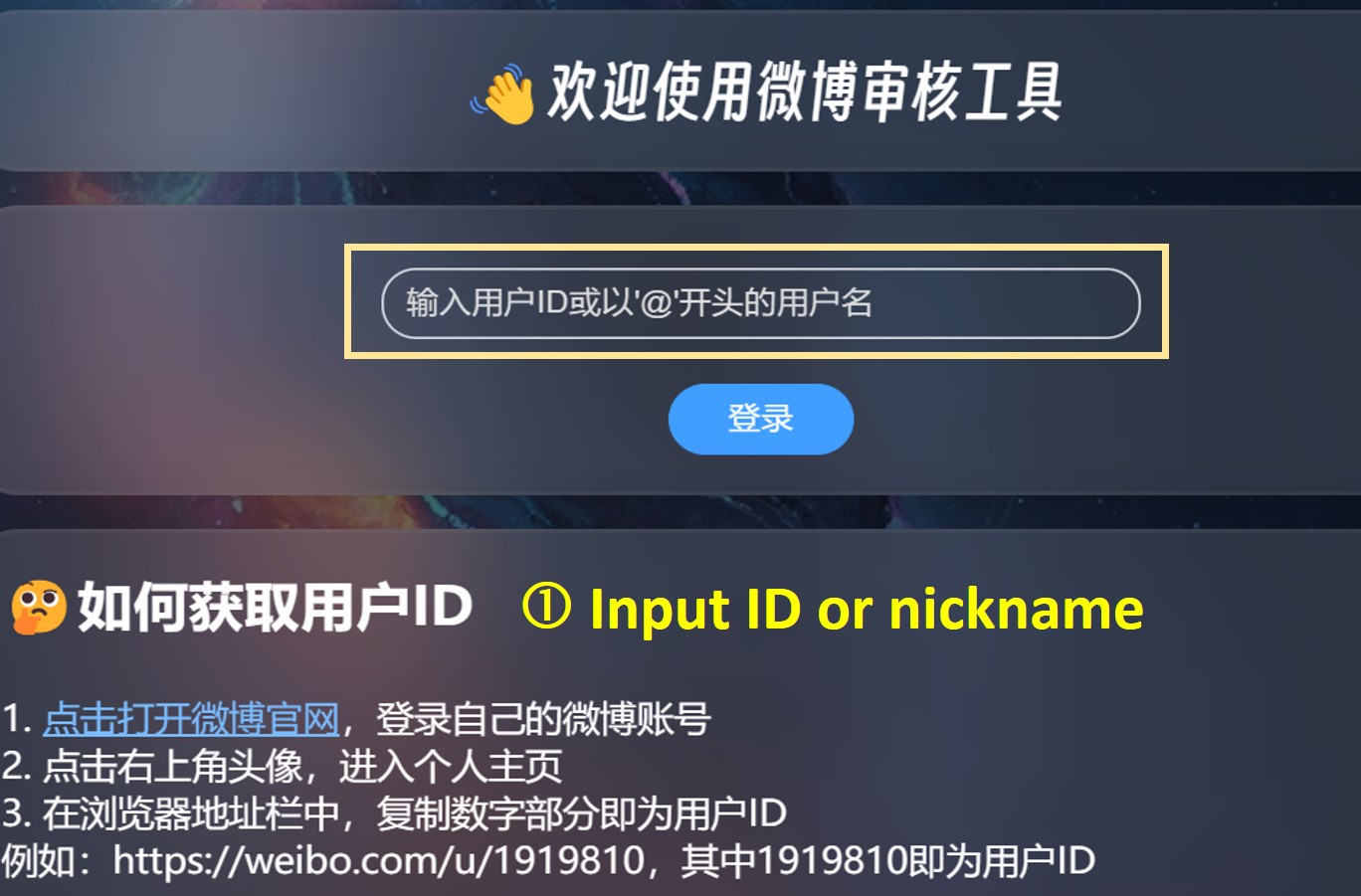}}
\hspace{0.001\linewidth}
\subfigure[Agree authorization]{\label{fig:authorization}
\includegraphics[width=0.31\linewidth,height=0.22\linewidth]{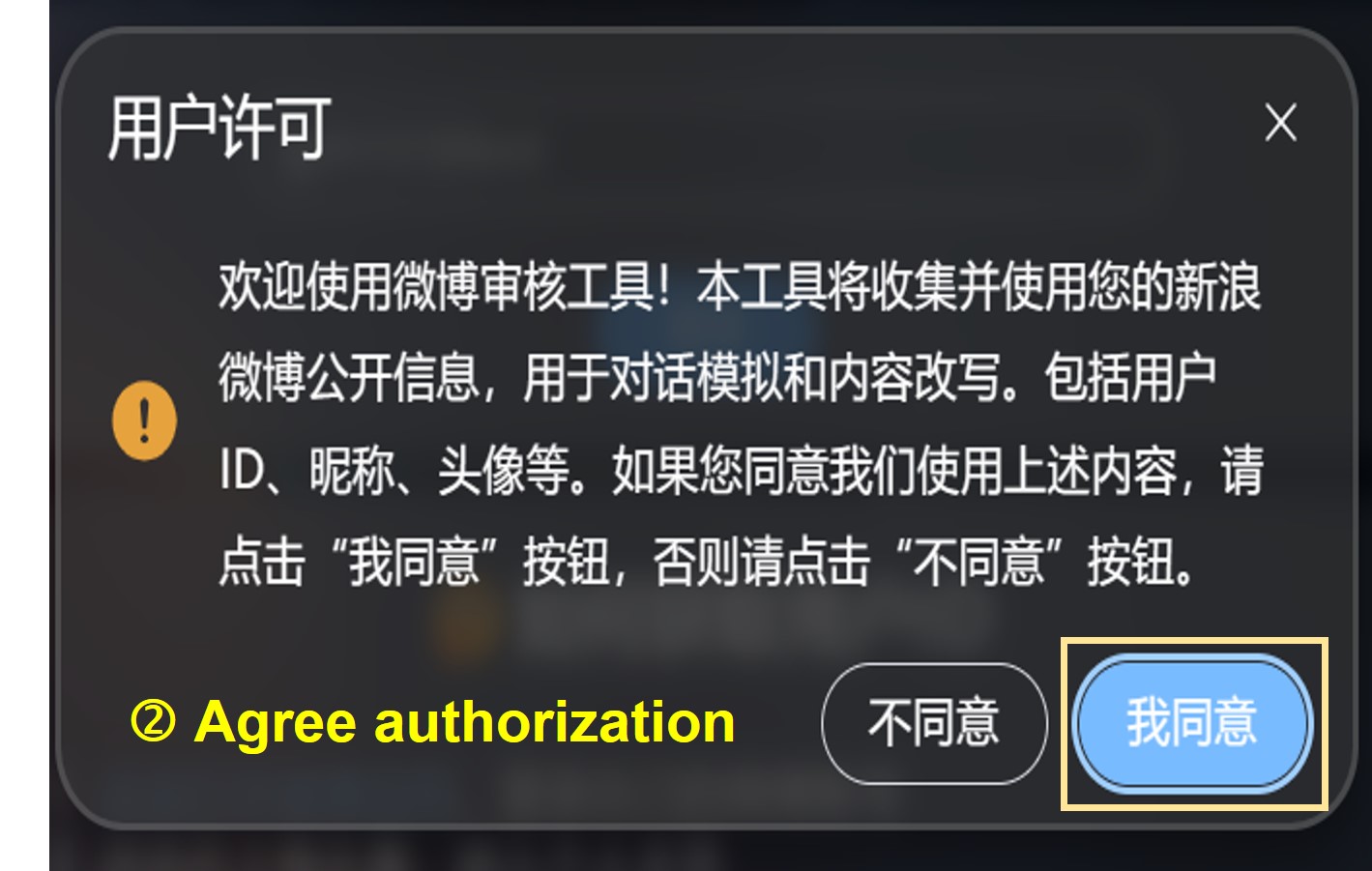}}
\hspace{0.001\linewidth}
\subfigure[Initial interface]{\label{fig:initial interface}
\includegraphics[width=0.31\linewidth,height=0.22\linewidth]{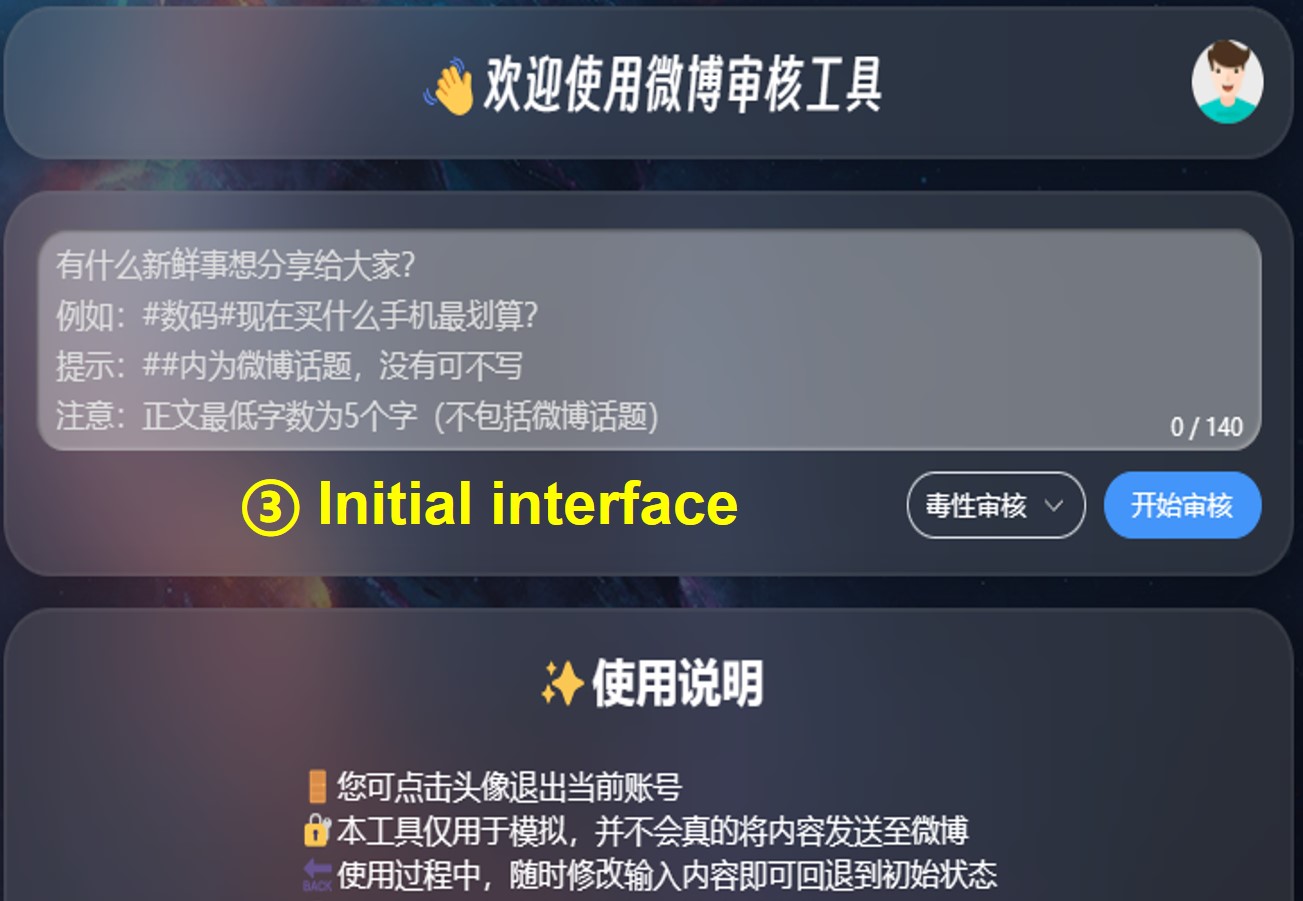}}
\caption{User login and authorization.}
\end{figure*}

A user can login into DeMod using her/his Weibo ID or "@nickname". Then there is an authorization statement, informing the user about the information it will access and use. When she/he agrees, DeMod presents the interface as depicted in Figure \ref{fig:initial interface}. The initial interface incorporates a text box, function buttons, and usage instructions. The background of the text box gives a description of the input format, including topic writing (use two "\#" signs to label the topic) and text length (the text should be at least five words without the topic). Users can input the text in the box and click the "Start" button for detection.

\begin{figure*}[h]
\centering
\subfigure[Toxicity detection]{\label{fig:toxicity detection}
\includegraphics[width=0.31\linewidth,height=0.22\linewidth]{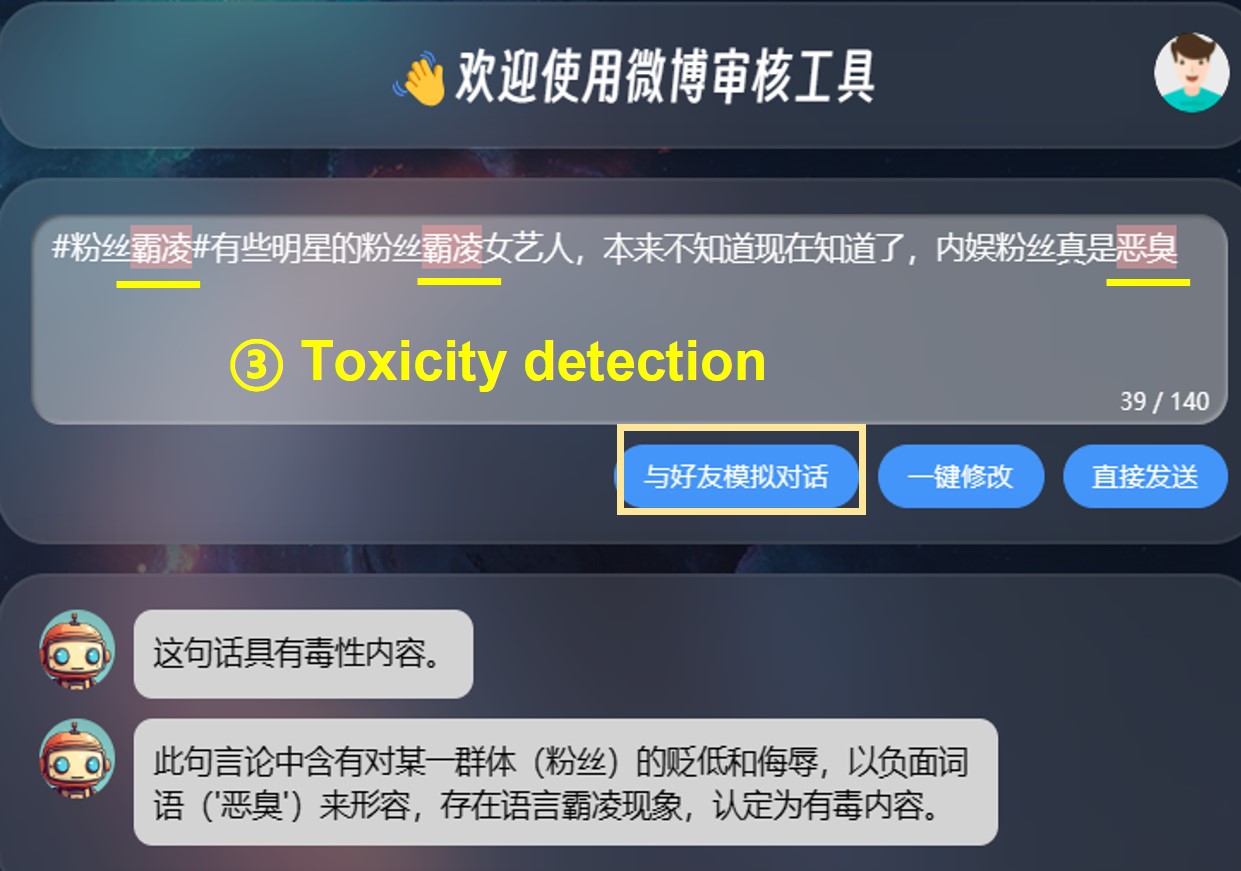}}
\hspace{0.001\linewidth}
\subfigure[Social role selection]{\label{fig:role selection}
\includegraphics[width=0.31\linewidth,height=0.22\linewidth]{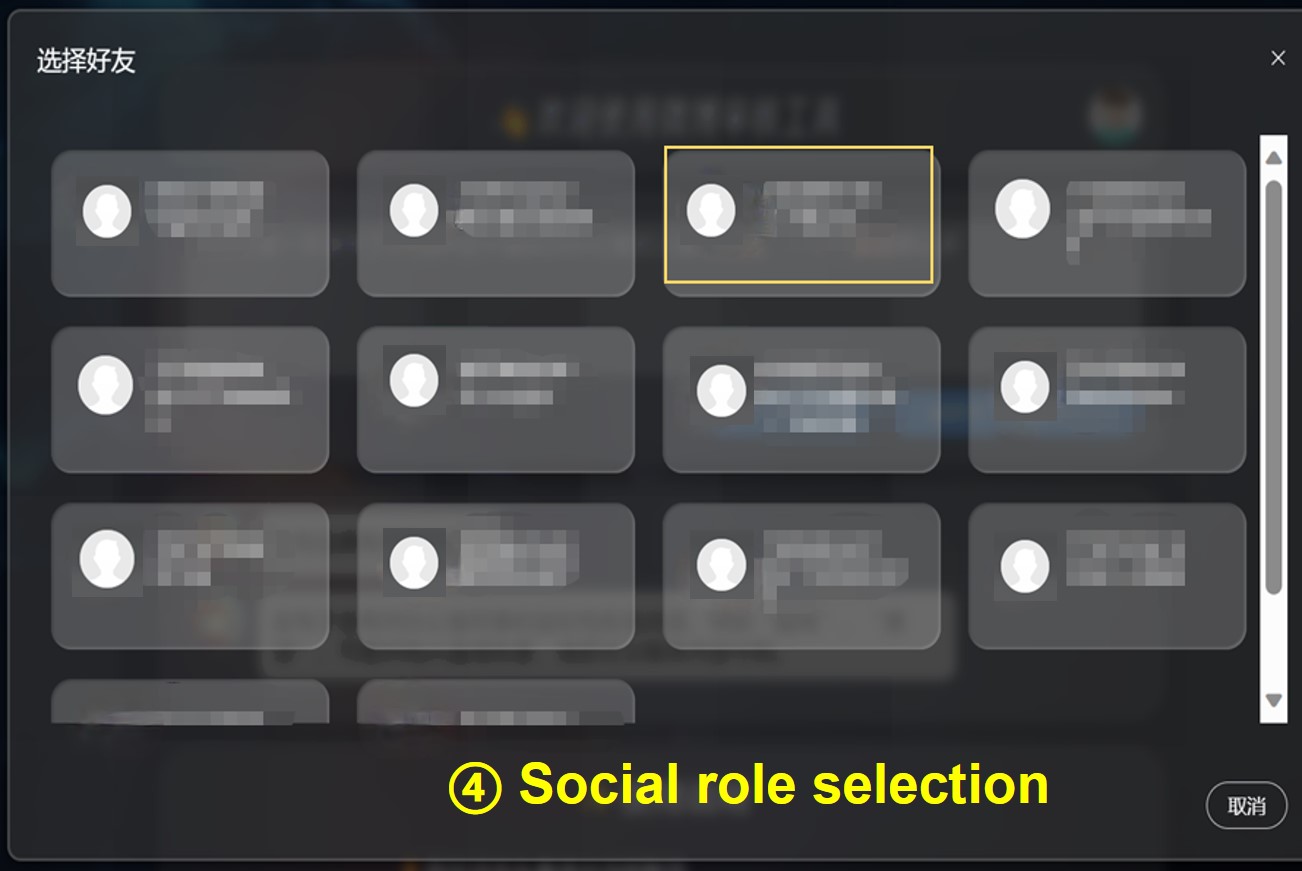}}
\vfill
\subfigure[Viewpoint simulation]{\label{fig:simulation}
\includegraphics[width=0.31\linewidth,height=0.22\linewidth]{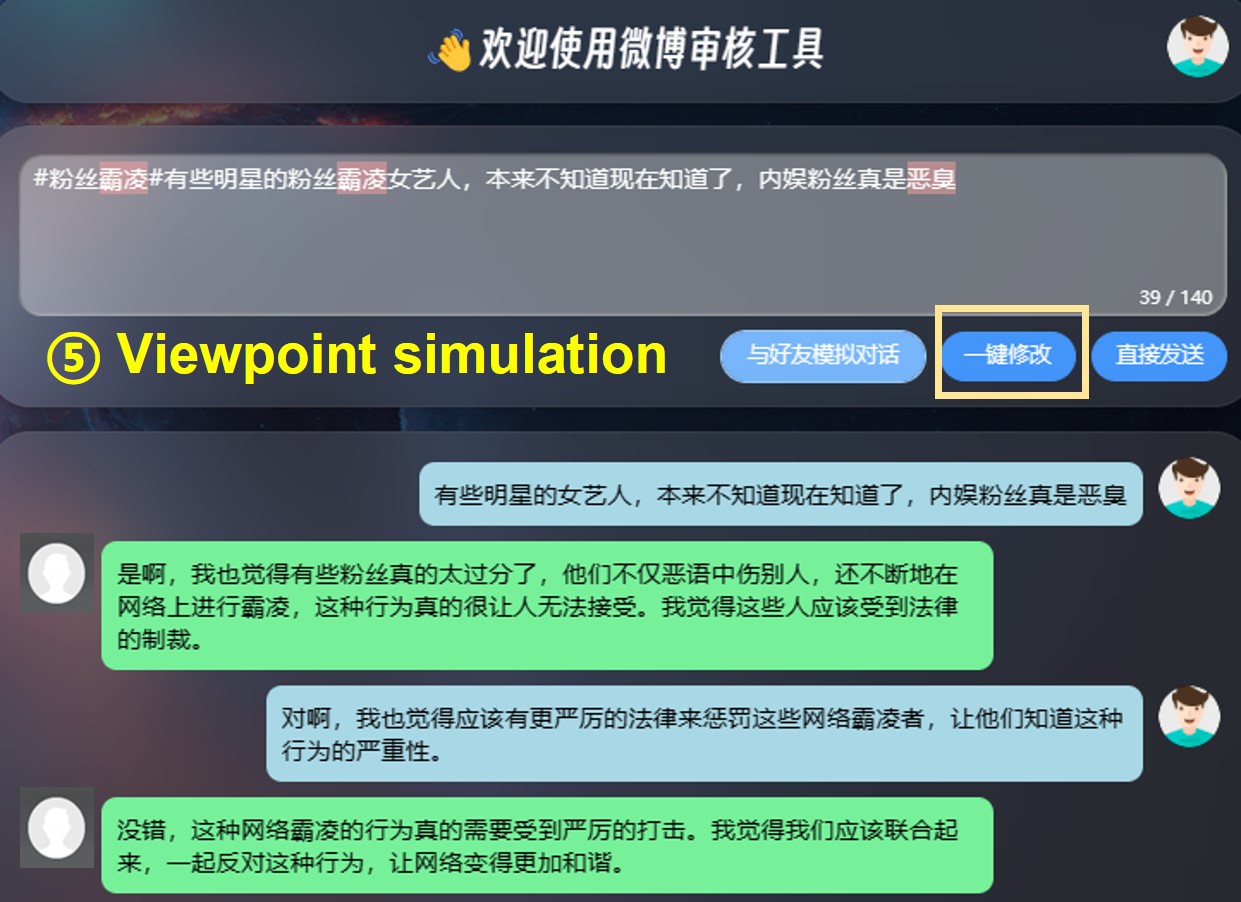}}
\hspace{0.001\linewidth}
\subfigure[Expression modification]{\label{fig:modification test}
\includegraphics[width=0.31\linewidth,height=0.22\linewidth]{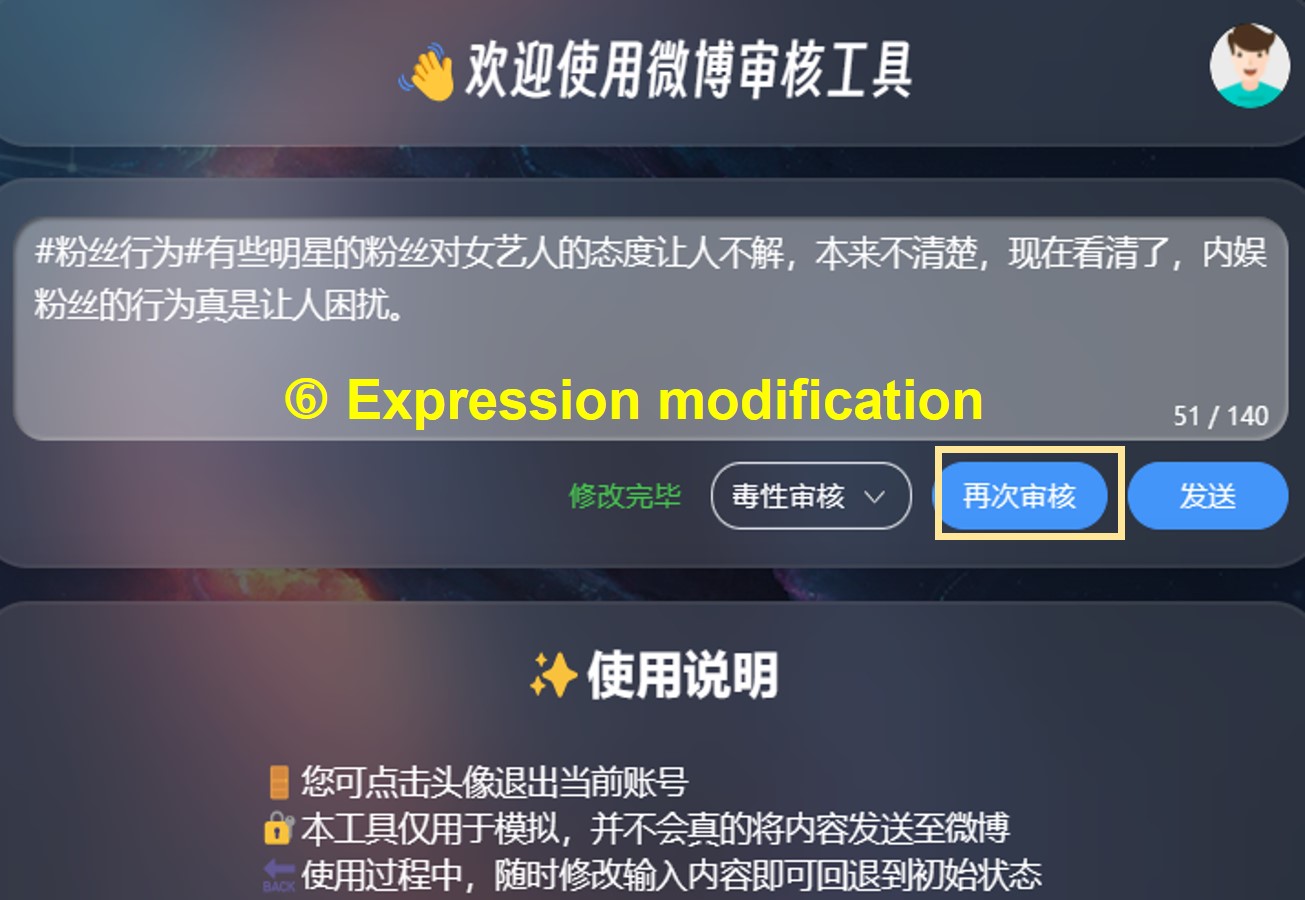}}
\caption{Toxicity censorship flow.}
\end{figure*}

We give an example of toxicity censorship, as shown in Figure \ref{fig:toxicity detection}. If the input is "\emph{\#FanBullying\# Some fans of celebrities bully female artists. I didn't know before, but now I do. The fans are really repulsive}" (This text is for demonstration only, and it is not an actual post), the detection results are displayed on the top left picture. The Y/N result indicates "\emph{This sentence contains toxic content}" and keywords are highlighted: "\emph{bully}" and "\emph{repulsive}". The immediate explanation is "\emph{This statement contains derogatory and insulting remarks towards a specific group (fans), employing negative words ('repulsive'). It indicates toxic behavior and language bullying}". If the user wants to get the audience's attitude to this post through simulation, click the "Simulate conversation with others" button and select the expected role in Figure \ref{fig:role selection}. The corresponding result is displayed in Figure \ref{fig:simulation}. If the user wants to revise the post, click the "Modify" button directly into the personalized modification. The modification result is "\emph{The attitude of some celebrities' fans towards female artists is perplexing. I didn't know before, but now I do. The fans are truly troubling}", shown in Figure \ref{fig:modification test}. Notably, the toxicity has been significantly reduced. If the user is dissatisfied or wishes to re-censor, click the "Re-censor" button and continue. The user can directly click "Send" to terminate. Here, "Send" doesn’t mean sharing with Weibo but synchronizing the content to the Weibo editing box for publishing. Besides, if users want to exit the current process or censor other content, re-edit in the text box. Users can click on their avatars in the upper-right corner to log out.

\section{Evaluation}
\subsection{Settings}
Around the design goals presented in Section 3, we conduct extensive evaluations for DeMod by employing several metrics, as shown in Table \ref{tab:evaluation dimensions}. For easy following, we organized these metrics into three dimensions, including function, performance, and design. The \textbf{function} dimension adopts integrity as the metric to measure to what extent DeMod meets users' function demands. For example, the "holistic integrity" means whether DeMod provides sufficient functions to support users' toxicity censorship, and the "explanation integrity" denotes whether the two kinds of explanations are sufficient. The \textbf{performance} dimension adopts accuracy to evaluate the effectiveness of DeMod, like "detection accuracy", "explanation accuracy", and "modification accuracy". The "detection accuracy" and "modification accuracy" are evaluated through both automatic and human evaluations, while "explanation accuracy", "semantic retention", and "personalized degree" can be assessed only through human evaluation. The \textbf{design} dimension involves two metrics: "ease of use" and "controllability", wherein the former reflects if DeMod is easy to follow, and the latter means whether users have sufficient control when using DeMod. Besides these dimensions, we also considered an \textbf{overall} dimension, including "user acceptance" and "strengths \& weaknesses" to reflect DeMod's overall user experience. The "user acceptance" measures if users would like to accept DeMod's censorship results, including the detection decision, explanation, and modification, and the "strengths \& weaknesses" reveals DeMod's overall strengths and weaknesses.

\begin{table}[!ht]
\centering
\small
\caption{Evaluation metrics and methods.}
\begin{tabular}{@{}cllcl@{}}
\hline
\textbf{Design goal} & \textbf{Description} & \textbf{Metric} & \textbf{Method}  & \textbf{Dimension} \\ \hline
G1 &  Provide holistic censorship  & Holistic integrity   & H  & Function  \\ \hline
\multirow{2}{*}{G2} &  \multirow{2}{*}{Offer fine-grained detection results} & Granularity integrity & H & Function   \\
~ & ~ & Detection accuracy  & A \& H  & Performance   \\ \hline
\multirow{2}{*}{G3} &  \multirow{2}{*}{Strengthen interpretability} & Explanation integrity  & H  & Function   \\
~ & ~ & Explanation accuracy  & H  & Performance   \\ \hline
\multirow{4}{*}{G4} &  \multirow{4}{*}{Give personalized revising suggestions} & Modification integrity  & H  & Function   \\
~ & ~ & Modification Accuracy  & A \& H  & Performance   \\
~ & ~ & Semantic retention   & H  & Performance   \\
~ & ~ & Personalized degree  & H  &  Performance  \\ \hline
\multirow{2}{*}{G5} &  \multirow{2}{*}{Ensure user-control} & Ease of use  & H  & Design   \\
~ & ~ & Controllability  & H  & Design   \\ \hline 
\end{tabular}
\\
``H'' indicates human evaluation and ``A'' indicates automatic evaluation.
\label{tab:evaluation dimensions}
\end{table}

A Chinese offensive language dataset named COLD \cite{deng2022cold_138} from Weibo and several baselines were employed for our automatic evaluations. The details of the dataset and baselines are given below.
\begin{itemize}
    \item \textbf{Dataset}. We utilized the COLD validation dataset \footnote{https://github.com/thu-coai/COLDataset} as our test dataset, comprising a total of 6,431 samples (3,211 toxic and 3,220 nontoxic posts). These 6,431 samples were employed in the detection task, and 3,211 toxic samples were used as the corpus of modification task, observing the effect of toxicity removal.
    \item \textbf{Baselines}. We utilized the Perspective API \cite{Perspective_140} and different versions of ChatGPT models as baselines, including GPT-3.5-turbo and GPT-4 \cite{OpenAI2023GPT4TR_99}. Perspective API is a commonly used automated tool for toxicity detection. It evaluates a toxic score (from 0 to 1) for the input text, wherein a higher score indicates stronger toxicity in the content. Referring to prior research \cite{Perspective_140}, we set 0.7 as the threshold, i.e., if the score returned by Perspective API is larger than 0.7, the content is toxic; nontoxic otherwise.     
\end{itemize}

To support human evaluation, we recruited participants to use DeMod in practice and give feedback. The details are as follows.
\begin{itemize}
    \item \textbf{Participants}. 28 interview participants described in Section 3 would like to further participate in our evaluation. We also posted a recruitment to attract new participants with the same requirements as our needfinding study. The introducing of new participants help improve the generalizability of our evaluation. Finally, 35 Weibo users participated in our evaluation (20 males and 15 females, aged 18 to 35).         
    \item \textbf{Procedure}. Human evaluation was conducted in one week, wherein the 35 participants freely chose to utilize DeMod for toxicity censorship without restriction. Only if participants have problems would we get involved. Users' operations were recorded into a log to support our analysis. In the meanwhile, we invited participants to fill out a questionnaire to give their feedback after one week's use. The questionnaire was designed with a 5-point Likert scale corresponding to each of the above metrics. The strengths \& weaknesses were reflected via two open-ended questions: "What do you think are the advantages of DeMod? " and "What do you think DeMod should improve? ".  
\end{itemize}

\subsection{Results}
For easy understanding, we present evaluation results in terms of the modules of DeMod. From the user logs, we found that participants used DeMod frequently. Specifically, each participant conducted toxicity detection for 7.029 times and modification for 4.543 times on average. For the questionnaire, we utilized Cronbach's coefficient alpha \cite{cho2015cronbach_161} to analyze the reliability of participants' feedback and got a result ${\alpha}=0.925$, indicating high reliability. Based on automatic evaluation and analyzing participants' logs and questionnaire responses, we obtained the following major findings. 

\subsubsection{Explainable Detection}
According to Table \ref{tab:evaluation dimensions}, both automatic and human evaluations were employed for the detection module. We first describe the accuracy of toxicity detection and then specify the other metrics reflected by our human evaluation.

In toxicity detection, DeMod outperforms the Perspective API significantly. DeMod with GPT-4 model achieves outstanding performance with accuracy reaching 73.50\%, and DeMod with GPT-3.5-turbo model gets an accuracy of 69.35\%, while the accuracy of Perspective API is 52.45\%. The results also inspire us to adopt GPT-4 as the core model to assist the implementation of DeMod.

As suggested in Table \ref{tab:evaluation dimensions}, the human evaluation in terms of the detection encompasses multiple dimensions, including function, performance, design, and overall assessment. Specifically, participants' evaluation results are shown in Figure \ref{fig:detect_a}, wherein the blue color represents a score of 4 or 5, the yellow represents 1 or 2, and the gray means 3. From the figure, we can see most participants appreciate the detection capability of DeMod, with an average score of over 4.1 in user acceptance. There were 32 participants (92.00\%) choosing 4 (willing) or 5 (very willing) in terms of acceptance of explainable detection. Over 33 participants (94.00\%) chose 4 (accurate) or 5 (very accurate) in terms of detection accuracy, with an average of 4.3. However, only 16 participants (45.00\%) selected 4 or 5 points regarding explanation accuracy, indicating it still needs improvement. Moreover, to study what factors affect user acceptance, we also explored the relationship between this dimension and the others. The result of chi-square test is shown in Table \ref{tab: Chi-squared acceptance}, where "$\ast$" indicates $p<0.05$, and "$\ast\ast$" means $p<0.01$. It can be found that user acceptance is positively correlated with several factors, including fine-grained integrity, explanation integrity, and detection accuracy. Additionally, from participants' open-ended responses, we further understood the above results and summarized the strengths and weaknesses of the detection module. 
\begin{itemize}
    \item \textbf{Strengths.} Participants respond that the DeMod's detector can help them identify post problems quickly and precisely, with descriptions like "\emph{an accurate detection}", "\emph{detailed explanations}", and "\emph{precise identifications}". Besides, participants think the dynamic explanation is a creative and novel design. The words like "\emph{novel}", "\emph{interesting}", "\emph{unique}", and "\emph{intriguing}" frequently appeared in participants' replies. P33 also mentioned, "\emph{The function is very great and quickly made me realize I shouldn't express so much emotion in my speech}". P4, P6, and P9 all expressed that this function was helpful for persons without good social interaction skills, saying "\emph{It's a little similar to the process of predicting the development trend of dialogue, which is necessary for users who suffer social phobia}".
    \item \textbf{Weaknesses.} Participants think the quality of the explanations could be enhanced, especially the dynamic explanation. The dynamic explanation needs to improve user experience, including user engagement and expected role's expression style. P4 suggested, "\emph{The response doesn't seem like my friend and it's not useful for me. Also, I can't continue to join the dialogue and express myself}". P30 also said, "\emph{Supporting interaction will be better}". Some participants offered some suggestions for DeMod. P12 suggested, "\emph{Could different levels be introduced in detection? Like discrimination, offensive language, insults and irony}". These suggestions will contribute to further improvement.
\end{itemize}

\subsubsection{Personalized Modification}
Both automatic and human evaluations were conducted for modification measurement according to Table \ref{tab:evaluation dimensions}. We first present the accuracy of toxicity modification, then specify the other dimensions reflected by human evaluation.

In toxicity modification, after DeMod's modification, the proportion of toxic samples has decreased by 94.38\%, from 3,211 to 170. It indicates DeMod's capability to revise toxic content. We also performed an analysis of the failed samples and found that these samples concentrate on the discrimination of region and gender. For example, "\emph{Shanghai always discriminates people from other places}" was modified to "\emph{Shanghai has some biases against from other places}", which is still identified as toxic. This phenomenon suggests the challenge of detoxifying.

\begin{figure*}
\centering
\subfigure[Detection]{\label{fig:detect_a}
\includegraphics[width=0.31\linewidth,height=0.22\linewidth]{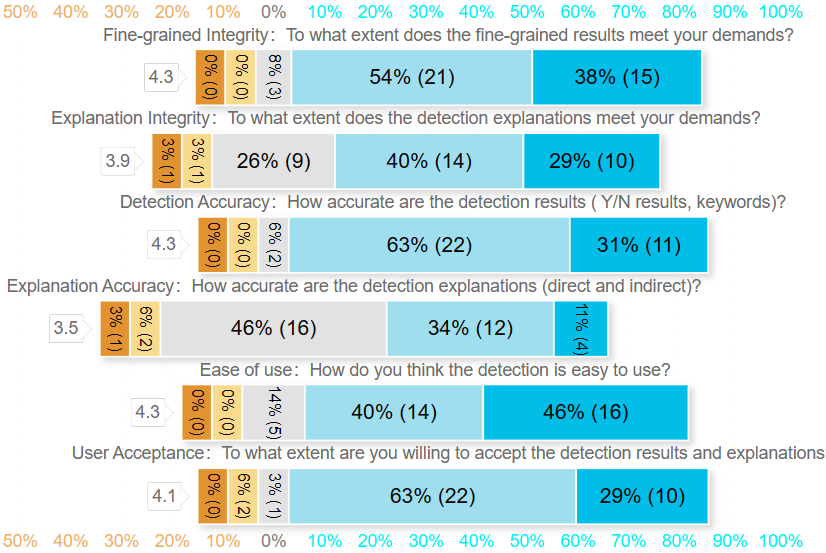}}
\hspace{0.01\linewidth}
\subfigure[Modification]{\label{fig:modify_b}
\includegraphics[width=0.3\linewidth,height=0.22\linewidth]{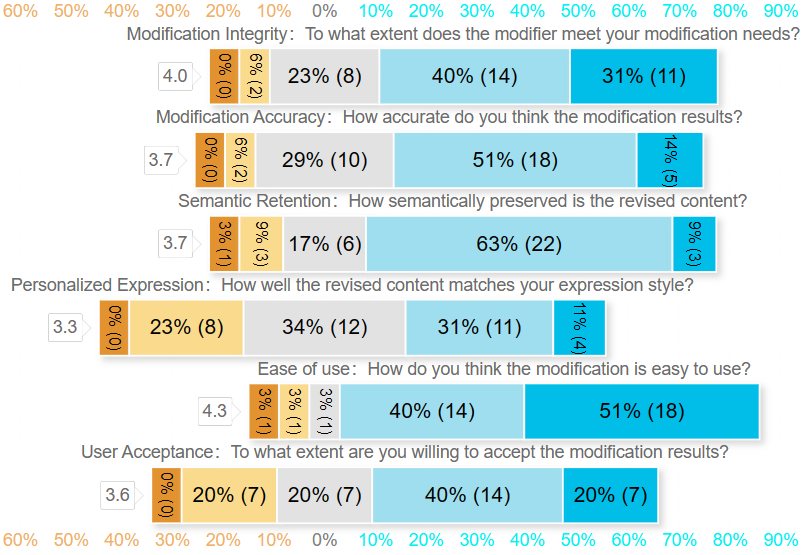}}
\hspace{0.01\linewidth}
\subfigure[DeMod]{\label{fig:overall_c}
\includegraphics[width=0.31\linewidth,height=0.22\linewidth]{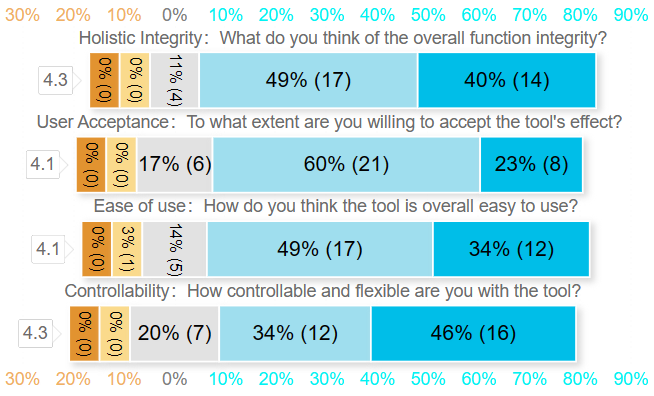}}
\label{fig:statistic}
\caption{Statistical visualization of human evaluation results.}
\end{figure*}

The human evaluation of modification encompasses function, performance, design, and overall assessment. Specifically, participants' evaluation results are shown in Figure \ref{fig:modify_b}, wherein the blue color represents a score of 4 or 5, the yellow represents 1 or 2, and the gray means 3. Participants overall expressed positive attitudes toward the modification's function integrity and ease of use, with an average of over 4.0 points. Over 65.00\% of the participants chose 4 or 5 regarding the modification accuracy and original semantic retention. Moreover, only 7 participants (20.00\%) explicitly responded that they were unwilling to accept the modified posts. Similar to the previous detection analysis, we conducted a chi-squared test to analyze the correlation between user acceptance of modification results and other metrics, as detailed in Table \ref{tab: Chi-squared acceptance}. The results suggest that user acceptance of modification is correlated with modification integrity, modification accuracy, and consistency with personalized expression style. Based on participants' open-ended feedback, we further understood the above results and summarized the strengths and weaknesses as follows.
\begin{itemize}
    \item \textbf{Strengths.} Several participants, such as P4, P7, P24, and P30, acknowledge the modification module's effectiveness in post revision. For example, P7 mentioned that "\emph{It is capable of modifying the offensive and insulting words automatically, and the modified text preserved the semantics as much as possible}". Besides, participants also appreciate the convenience of the modification module. P10 mentioned, "\emph{The one-click button for automated modification is very convenient. The modification overall meets my needs and posts can be published with little change}". Responses from participants like P2, P8, P9, P21, P28, and P31 included terms like "\emph{convenient}", "\emph{easy}", and "\emph{automatic}".     
    \item \textbf{Weaknesses.} However, some participants feel that the emotion changes a lot before and after the modification, saying "\emph{The toxicity modification was quite problematic, often completely altering the original intent and changing a critical attitude to a neutral or even positive one}". A few users also suggest adding functionality to compare the results before and after modification, saying "\emph{I hope it can show the original content, helping me quickly see what has been modified}".
\end{itemize}

\begin{table}[!ht]
    \centering
    \small
    \caption{Chi-squared results between user acceptance and diverse metrics in modules.}
    \begin{tabular}{@{}cp{5cm}cc@{}}
    \hline
        \textbf{User Acceptance} & \makebox[0.2\textwidth][c]{\textbf{Metric}} & \textbf{${\chi}^2$} & \textbf{p} \\ \hline
        \multirow{5}*{Detection} & Function$-$Fine-grained integrity& 12.812 & 0.046$\ast$ \\ 
        ~ & Function$-$Explanation integrity & 30.64 & 0.002$\ast$ \\ 
        ~ & Performance$-$Detection accuracy& 17.167 & 0.009$\ast\ast$ \\ 
        ~ & Performance$-$Explanation accuracy& 15.505 & 0.215 \\ 
        ~ & Design$-$Ease of use of detection & 10.177 & 0.117 \\ \hline
        \multirow{5}*{Modification} & Function$-$Modification integrity & 21.23 & 0.012$\ast$ \\ 
        ~ & Performance$-$Modification accuracy & 24.111 & 0.004$\ast\ast$ \\ 
        ~ & Performance$-$Semantic retention & 18.788 & 0.094 \\ 
        ~ & Performance$-$Personalized expression & 17.32 & 0.044$\ast$ \\ 
        ~ & Design$-$Ease of use of modification & 11.25 & 0.508 \\ \hline
    \end{tabular}
    \label{tab: Chi-squared acceptance}
\end{table}

\subsubsection{Overall}
The result of the overall evaluation is shown in Figure \ref{fig:overall_c}. The average score is over 4.0 on each metric, including holistic integrity, user acceptance, ease of use, and controllability. More than 28 participants (80.00\%) believe that DeMod addresses their censorship needs, and 29 participants (83.00\%) suggest that the operation is convenient and easy to study. We also surveyed participants' preferences on DeMod's different functions. 26 participants (74.26\%) chose the detection, and 17 (48.57\%) preferred modification. To explore the relationship in the same dimension between overall DeMod and each module, we conducted a chi-squared test to analyze the correlation for all dimensions, as shown in Table \ref{tab:chi-squared overall}. The results suggest that the function, performance, and design of overall DeMod are correlated with each metric of modules. Based on participants' open-ended feedback, we further understood the above results and summarized DeMod's overall strengths and weaknesses as follows.

\begin{itemize}
    \item \textbf{Strengths.} Most participants suggest that DeMod is an innovative censorship tool that effectively meets their censorship needs. It can not only detect the potential problems of posts but also provide explanations and solutions. P4 mentioned, "\emph{The censorship tool provides solutions to revise content, rather than just telling me there are problems}". P16 stated, "\emph{There hasn't been a similar tool before, and it solves the end-users censorship challenges. It's entirely controllable and relatively easy to use}". P34 also emphasized, "\emph{When I am angry or depressed, the viewpoint simulation can make me a bit relaxed and more objective}". Additionally, 6 participants, such as P5 and P10, explicitly acknowledged DeMod's value in social interaction, highlighting its capability to "\emph{avoid social conflicts}", "\emph{make the social environment more normal}", etc.
    \item \textbf{Weaknesses.} Participants wish to improve the efficiency of DeMod's censorship process. Some participants pointed out that it was time-consuming when using DeMod for content censorship, especially viewpoint simulation. P11 replied, "\emph{The efficiency needs improvement, the dialogue generation is too slow}", and P13 stated, "\emph{I'm not sure if a user wants to spend time waiting for the simulation}".
\end{itemize}

\begin{table}[!ht]
    \centering
    \small
    \caption{Chi-squared results between overall and modules in dimensions.}
    \begin{tabular}{@{}cp{5cm}cc@{}}
    \hline
        \textbf{Dimension} & \makebox[0.2\textwidth][c]{\textbf{Metric}} & \textbf{${\chi}^2$} & \textbf{p} \\ \hline
        \multirow{3}*{Function} & Detection$-$Fine-grained integrity & 17.106 & 0.002$\ast\ast$ \\
        ~ & Detection$-$Explanation integrity & 24.19 & 0.002$\ast\ast$ \\ 
        ~ & Modification$-$Modification integrity & 15.002 & 0.020$\ast$ \\ \hline
        \multirow{5}*{Performance} & Detection$-$Detection accuracy & 11.326 & 0.023$\ast$ \\ 
        ~ & Detection$-$Explanation accuracy  & 20.707 & 0.008** \\ 
        ~ & Modification$-$Modification accuracy & 22.519 & 0.001$\ast\ast$ \\ 
        ~ & Modification$-$Semantic retention& 22.074 & 0.005$\ast\ast$ \\ 
        ~ & Modification$-$Personalized expression & 30.62 & 0.000$\ast\ast$ \\ \hline
        \multirow{2}*{Design} & Detection$-$Ease of use of detection & 30.823 & 0.000$\ast\ast$ \\ 
        ~ & Modification$-$Ease of use of modification & 32.684 & 0.001$\ast\ast$ \\ \hline
    \end{tabular}
    \label{tab:chi-squared overall}
\end{table}

\textbf{Summary.} To conclude, the above evaluations indicate DeMod's capability to support toxicity censorship and high acceptance among participants. The detector provides accurate and fine-grained detection results and different kinds of explanations, and the modifier supplies effective personalized modification suggestions, jointly promoting users' multi-stage procedure of identifying, understanding, and modifying in content censorship. However, the evaluation results also highlight the potential of DeMod's improvements in the future, including enhancing user engagement and expression style in dynamic explanations, promoting emotion consistency in content modification, and improving the efficiency of the whole framework. 

\section{Discussion}
In this paper, we investigate social media users' current toxicity censorship practices and gain several insights into the design of content censorship tools. We found that users had diverse demands for the design of toxicity censorship tools, including providing holistic censorship, offering fine-grained detection results, strengthening interpretability, giving personalized revising suggestions, and ensuring user-control, while existing approaches and tools mostly focus on toxicity detection. Therefore, we propose the novel holistic content censorship tool, DeMod. By taking advantage of ChatGPT, DeMod is equipped with capabilities of explainable detection and content modification, helping users conduct toxicity censorship more comprehensively, efficiently, and flexibly. Evaluations reveal that DeMod is widely used and accepted by participants. Its integrated features, including accurate and explainable detection and personalized modification suggestions, have significantly improved users' ability to identify and modify content during the censorship process.

\subsection{Implications}
Drawing on the needfinding study (Sec 3) and our evaluations (Sec 5), we present the following implications for future research and design of toxicity censorship systems.

\textbf{Promoting content censorship from the holistic perspective.} One crucial finding from our needfinding study (Sec 3.2) and DeMod's evaluation results (Sec 5.2) concentrates on social media users' diverse and complicated demands on the functions of toxicity censorship tools, beyond just toxicity detection. Although prior research has employed methods to help users identify toxic content on social media, such as Google's Perspective \cite{Perspective_140} and RECAST \cite{Wright_48}, these methods and tools focus on supporting toxicity detection without further processing. Our needfinding study and evaluation results confirm users' detection needs but suggest more expectations. Not only does the granularity of detection and immediate explanation need to be enhanced, but also some new functions including dynamic explanation and modification recommendation should be introduced in designing a holistic toxicity censorship tool (Sec 3.2). In this context, DeMod is designed to be equipped with multiple functions, including fine-grained accurate detection, immediate and dynamic explanations, and personalized modification suggestions. The design of DeMod has enriched the functions of content censorship and refined the granularity, promoting efficiency and user experience of toxicity censorship. Participants in our evaluations explicitly expressed their appreciation of DeMod's capabilities and willingness to utilize its multiple functions (Sec 5.2).

\textbf{Emphasizing the interpretability of censorship process and results.} This insight highlights the significance of providing different kinds of explanations along with identification results to promote user understanding (Sec 3.2). Enhancing user understanding is an essential task in the research of toxicity censorship and content moderation. Previous studies have addressed this problem mainly from the perspective of refining the granularity of identification results and highlighting the fine-grained results. \cite{Wright_48, Liu_84} considered this thought and designed toxicity and sensitive information detection tools with features like keyword identification, highlighting, bolding, and ranking. 
Although these features can enhance user understanding to some extent, such methods still lack explanations, e.g., why these keywords are toxic content and what the influence of such content will be, especially when a user cannot comprehend some keywords' meanings well. For this problem, we offer two new kinds of explanations besides keyword identification and highlighting - immediate and dynamic explanations (Sec 4.1.2). 
The dynamic explanation module is a novel design based on LLMs' simulation capability, leading to a user's deeper understanding of the potential consequences and consciousness of toxic content sharing. This function can encourage users to pay attention to responsible disclosure on social media platforms, and further regulate their social interaction behaviors.

\textbf{Emphasizing the balance of multiple targets in modification.} DeMod's modification procedure is designed to achieve multiple targets, including automatic detoxifying, original semantics retention, and personalized language style integration. However, achieving a balance of these goals requires further improvement. In our evaluations, users expressed concerns about the accuracy of detoxifying. They worry that detoxifying effectively can lead to a loss of the original expression intent, thereby affecting their authenticity on social platforms. These concerns are understandable since overly aggressive modifications can make content stiff and unnatural, significantly diverging from the user's authentic expression (Sec 5.2.2). Therefore, it is essential to find a balance, offering detoxifying while respecting the expression intent of users. How to investigate this balance and master it in toxicity censorship is a promising topic in future research.

\textbf{Ensuring user-control.} In our needfinding study and evaluations, users have emphasized the demand for engaging and controlling their content regulation procedure (Sec 3.2 and Sec 5.2). However, social media platforms currently have not provided functions to help users conduct toxicity censorship while relying on measures like post-moderation, content deletion, or account suspension \cite{Zhao_86}, making users passive and losing control of content regulation (Sec 3.2). Compared with that, DeMod pays attention to users' perception, understanding, and control of toxic content censorship (Sec 5.2), e.g., perceiving the fine-grained detection results, understanding the rationale, and controlling the modification. Moreover, in the design of DeMod, we split the framework into several independent functional modules, and users can use different functions easily. Each module is characterized by solid internal consistency and specifications, resulting in the loose coupling between modules and each module's flexible usage and promotion. 

\subsection{Impacts for Other Stakeholders}
DeMod is a user-centered automated tool to help social media users conduct toxic content censorship. Besides social media users, social media platform practitioners and policymakers can also benefit from this tool. From the perspective of platforms, integrating such content censorship functions can proactively prevent toxic content, reducing their efforts and costs on content moderation. From the perspective of policymakers, DeMod provides insights into promoting content regulation policies. As a real application, DeMod exhibits how advanced techniques and novel design can effectively address content censorship problems in the social media context. Policymakers can draw inspiration from DeMod's practice to encourage and support more technical innovations in response to the evolving ethical and legal challenges in social environments. 

\subsection{Ethical Considerations}
A potential concern is there might be misuse of DeMod's functions. Although DeMod is designed to enhance users' censorship practices and improve the quality of social interaction, techniques are often associated with potential risks of misuse. Malicious users might leverage DeMod's explainable detector to craft a speech to mislead others or incite conflicts. To prevent such problems, practitioners can adopt several measures. First, strengthening user authentication ensures that only normal users can use DeMod's features. Second, setting a threshold to limit the times of viewpoint simulation. Third, automated monitoring and review mechanisms should be introduced to identify and address misuse behaviors in time. Utilizing these strategies helps prevent and intervene in the misuse of DeMod and ensures that DeMod provides effective service without damage.

\section{Limitations and Future Work}
As the first work addressing the holistic censorship of toxicity, this research suffers from some limitations. The first one lies in DeMod's performance. Although the effectiveness of DeMod is acceptable, it still sometimes lacks accuracy in detection and modification. Specifically, the dynamic explanation module is occasionally unable to understand the context well and give accurate explanations. We thought there are two reasons for this problem. The first is that some social posts are expressed using buzzwords or jargon, and the LLMs cannot comprehend these words' semantics, intent, and sentiment very well. With the increase of training corpus and fine-tuning techniques, LLMs' capability in understanding the specific or domain words can be improved, bringing opportunities to alleviate DeMod's accuracy problems. The second reason is related to the prompt design. Although we have tried various prompts and selected the most effective ones, their effectiveness cannot be guaranteed to support all scenarios. Prompt engineering is essentially a complicated task since LLMs remain a ``black-box''. In the future, we plan to optimize DeMod to fully meet users' expectations for censorship performance through conducting context-sensitive prompt engineering methods and diverse feature exploration.

The second limitation is that DeMod only employs a text-based LLM - ChatGPT as a backbone to process toxic content, while social posts usually contain multimedia content such as images or videos. The post images and videos can also have some toxic content. Therefore, incorporating multi-modal data modeling and processing techniques into DeMod for multimedia content censorship is a promising research topic in the future. We can see there emerge several multi-modal pre-trained large models (PLMs) like Llama2 \cite{touvron2023llama_95} and LLaVA \cite{liu2023visual_96}. 
Since DeMod is designed in a modular manner, these diverse models can be easily introduced into it (replacing ChatGPT with the other open-source or closed-source multi-modal LLMs and invoking the corresponding APIs) to enhance its performance in identifying multi-modal toxicity \cite{de2024rtp_182}.

The last limitation is that this study's needfinding study and evaluations were conducted just by using Weibo as a research site, which might limit our findings' generalizability. Different social media platforms differ from each other in user characteristics, cultures, and openness. The differences can influence users’ content censorship demands and hinder DeMod’s general use across different platforms. As the first work on studying a holistic toxicity censorship tool, we focus on the crucial modules of system design. It is better to validate DeMod's performance variance across different platforms and design adaptive strategies in the future, which requires collaborating with sociology experts to gain more insights into people's toxicity censorship practices in different social contexts. 
Additionally, DeMod is just implemented as a prototype in our work. Future work on system design can explore its real large-scale deployment by combining with frameworks like federated learning \cite{banabilah2022federated_183}. Based on that, long-term observations can be conducted, and a more comprehensive understanding of its application, potential problems, and improvements will be achieved.

\section{Conclusion}
In this work, we develop DeMod, a holistic toxicity censorship tool, incorporating features of explainable \textbf{De}tection and personalized \textbf{Mod}ification. Through different kinds of explanations and modification recommendations, DeMod reduces users' censorship loads and improves their experience. Extensive evaluations suggest DeMod's multiple strengths like the richness of functionality, the performance of censorship, and ease of use. Our results also lead to several innovative insights into the future censorship system research and design, promoting the building of friendly online communities. In the future, we will focus on improving DeMod's capability for multimedia content censorship and promoting its application in more scenarios.

\bibliographystyle{ACM-Reference-Format}
\bibliography{DeMod-authordraft}

%%% -*-BibTeX-*-
%%% Do NOT edit. File created by BibTeX with style
%%% ACM-Reference-Format-Journals [18-Jan-2012].

\begin{thebibliography}{66}

%%% ====================================================================
%%% NOTE TO THE USER: you can override these defaults by providing
%%% customized versions of any of these macros before the \bibliography
%%% command.  Each of them MUST provide its own final punctuation,
%%% except for \shownote{}, \showDOI{}, and \showURL{}.  The latter two
%%% do not use final punctuation, in order to avoid confusing it with
%%% the Web address.
%%%
%%% To suppress output of a particular field, define its macro to expand
%%% to an empty string, or better, \unskip, like this:
%%%
%%% \newcommand{\showDOI}[1]{\unskip}   % LaTeX syntax
%%%
%%% \def \showDOI #1{\unskip}           % plain TeX syntax
%%%
%%% ====================================================================

\ifx \showCODEN    \undefined \def \showCODEN     #1{\unskip}     \fi
\ifx \showDOI      \undefined \def \showDOI       #1{#1}\fi
\ifx \showISBNx    \undefined \def \showISBNx     #1{\unskip}     \fi
\ifx \showISBNxiii \undefined \def \showISBNxiii  #1{\unskip}     \fi
\ifx \showISSN     \undefined \def \showISSN      #1{\unskip}     \fi
\ifx \showLCCN     \undefined \def \showLCCN      #1{\unskip}     \fi
\ifx \shownote     \undefined \def \shownote      #1{#1}          \fi
\ifx \showarticletitle \undefined \def \showarticletitle #1{#1}   \fi
\ifx \showURL      \undefined \def \showURL       {\relax}        \fi
% The following commands are used for tagged output and should be
% invisible to TeX
\providecommand\bibfield[2]{#2}
\providecommand\bibinfo[2]{#2}
\providecommand\natexlab[1]{#1}
\providecommand\showeprint[2][]{arXiv:#2}

\bibitem[Per(2023)]%
        {Perspective_140}
 \bibinfo{year}{2023}\natexlab{}.
\newblock \bibinfo{title}{Perspective API}.
\newblock
\newblock
\newblock
\shownote{\url{https://www.perspectiveapi.com/research/}}.


\bibitem[Aghajari et~al\mbox{.}(2023)]%
        {Aghajari_26}
\bibfield{author}{\bibinfo{person}{Zhila Aghajari}, \bibinfo{person}{Eric P.~S. Baumer}, {and} \bibinfo{person}{Dominic DiFranzo}.} \bibinfo{year}{2023}\natexlab{}.
\newblock \showarticletitle{What’s the Norm Around Here? Individuals’ Responses Can Mitigate the Effects of Misinformation Prevalence in Shaping Perceptions of a Community}. In \bibinfo{booktitle}{\emph{Proceedings of the 2023 CHI Conference on Human Factors in Computing Systems}}. \bibinfo{address}{New York, NY, USA}, Article \bibinfo{articleno}{550}, \bibinfo{numpages}{27}~pages.
\newblock


\bibitem[Ahmad et~al\mbox{.}(2020)]%
        {TangiblePrivacyUserCentric2020_178}
\bibfield{author}{\bibinfo{person}{Imtiaz Ahmad}, \bibinfo{person}{Rosta Farzan}, \bibinfo{person}{Apu Kapadia}, {and} \bibinfo{person}{Adam~J. Lee}.} \bibinfo{year}{2020}\natexlab{}.
\newblock \showarticletitle{Tangible Privacy: Towards User-Centric Sensor Designs for Bystander Privacy}.
\newblock \bibinfo{journal}{\emph{Proceedings of the ACM on Human-Computer Interaction}} \bibinfo{volume}{4}, \bibinfo{number}{CSCW2}, Article \bibinfo{articleno}{116} (\bibinfo{year}{2020}), \bibinfo{numpages}{28}~pages.
\newblock


\bibitem[Banabilah et~al\mbox{.}(2022)]%
        {banabilah2022federated_183}
\bibfield{author}{\bibinfo{person}{Syreen Banabilah}, \bibinfo{person}{Moayad Aloqaily}, \bibinfo{person}{Eitaa Alsayed}, \bibinfo{person}{Nida Malik}, {and} \bibinfo{person}{Yaser Jararweh}.} \bibinfo{year}{2022}\natexlab{}.
\newblock \showarticletitle{Federated Learning Review: Fundamentals, Enabling Technologies, and Future Applications}.
\newblock \bibinfo{journal}{\emph{Information Processing \& Management}} \bibinfo{volume}{59}, \bibinfo{number}{6} (\bibinfo{year}{2022}), \bibinfo{pages}{103061}.
\newblock
\showISSN{0306-4573}


\bibitem[Beres et~al\mbox{.}(2021)]%
        {Beres_34}
\bibfield{author}{\bibinfo{person}{Nicole~A Beres}, \bibinfo{person}{Julian Frommel}, \bibinfo{person}{Elizabeth Reid}, \bibinfo{person}{Regan~L Mandryk}, {and} \bibinfo{person}{Madison Klarkowski}.} \bibinfo{year}{2021}\natexlab{}.
\newblock \showarticletitle{Don’t You Know That You’re Toxic: Normalization of Toxicity in Online Gaming}. In \bibinfo{booktitle}{\emph{Proceedings of the 2021 CHI Conference on Human Factors in Computing Systems}}. \bibinfo{address}{New York, NY, USA}, Article \bibinfo{articleno}{438}, \bibinfo{numpages}{15}~pages.
\newblock


\bibitem[Blackwell et~al\mbox{.}(2018)]%
        {Blackwell_72}
\bibfield{author}{\bibinfo{person}{Lindsay Blackwell}, \bibinfo{person}{Tianying Chen}, \bibinfo{person}{Sarita Schoenebeck}, {and} \bibinfo{person}{Cliff Lampe}.} \bibinfo{year}{2018}\natexlab{}.
\newblock \showarticletitle{When Online Harassment is Perceived as Justified}. In \bibinfo{booktitle}{\emph{Proceedings of the 12th International AAAI Conference on Web and Social Media}}. \bibinfo{pages}{1--10}.
\newblock


\bibitem[Brown et~al\mbox{.}(2020)]%
        {Brown_92}
\bibfield{author}{\bibinfo{person}{Tom~B. Brown}, \bibinfo{person}{Benjamin Mann}, \bibinfo{person}{Nick Ryder}, \bibinfo{person}{Melanie Subbiah}, \bibinfo{person}{Jared Kaplan}, \bibinfo{person}{Prafulla Dhariwal}, \bibinfo{person}{Arvind Neelakantan}, \bibinfo{person}{Pranav Shyam}, \bibinfo{person}{Girish Sastry}, \bibinfo{person}{Amanda Askell}, \bibinfo{person}{Sandhini Agarwal}, \bibinfo{person}{Ariel Herbert-Voss}, \bibinfo{person}{Gretchen Krueger}, \bibinfo{person}{Tom Henighan}, \bibinfo{person}{Rewon Child}, \bibinfo{person}{Aditya Ramesh}, \bibinfo{person}{Daniel~M. Ziegler}, \bibinfo{person}{Jeffrey Wu}, \bibinfo{person}{Clemens Winter}, \bibinfo{person}{Christopher Hesse}, \bibinfo{person}{Mark Chen}, \bibinfo{person}{Eric Sigler}, \bibinfo{person}{Mateusz Litwin}, \bibinfo{person}{Scott Gray}, \bibinfo{person}{Benjamin Chess}, \bibinfo{person}{Jack Clark}, \bibinfo{person}{Christopher Berner}, \bibinfo{person}{Sam McCandlish}, \bibinfo{person}{Alec Radford}, \bibinfo{person}{Ilya Sutskever},
  {and} \bibinfo{person}{Dario Amodei}.} \bibinfo{year}{2020}\natexlab{}.
\newblock \showarticletitle{Language Models are Few-shot Learners}. In \bibinfo{booktitle}{\emph{Proceedings of the 34th International Conference on Neural Information Processing Systems}}. \bibinfo{address}{Red Hook, NY, USA}, Article \bibinfo{articleno}{159}, \bibinfo{numpages}{25}~pages.
\newblock


\bibitem[Cai and Wohn(2019)]%
        {Cai_68}
\bibfield{author}{\bibinfo{person}{Jie Cai} {and} \bibinfo{person}{Donghee~Yvette Wohn}.} \bibinfo{year}{2019}\natexlab{}.
\newblock \showarticletitle{What are Effective Strategies of Handling Harassment on Twitch? Users' Perspectives}. In \bibinfo{booktitle}{\emph{Companion Publication of the 2019 Conference on Computer Supported Cooperative Work and Social Computing}}. \bibinfo{address}{New York, NY, USA}, \bibinfo{pages}{166–170}.
\newblock


\bibitem[Cai and Wohn(2021)]%
        {Cai_5}
\bibfield{author}{\bibinfo{person}{Jie Cai} {and} \bibinfo{person}{Donghee~Yvette Wohn}.} \bibinfo{year}{2021}\natexlab{}.
\newblock \showarticletitle{After Violation But Before Sanction: Understanding Volunteer Moderators' Profiling Processes Toward Violators in Live Streaming Communities}.
\newblock \bibinfo{journal}{\emph{Proceedings of the ACM on Human-Computer Interaction}} \bibinfo{volume}{5}, \bibinfo{number}{CSCW2}, Article \bibinfo{articleno}{410} (\bibinfo{year}{2021}), \bibinfo{numpages}{25}~pages.
\newblock


\bibitem[Cho and Kim(2015)]%
        {cho2015cronbach_161}
\bibfield{author}{\bibinfo{person}{Eunseong Cho} {and} \bibinfo{person}{Seonghoon Kim}.} \bibinfo{year}{2015}\natexlab{}.
\newblock \showarticletitle{Cronbach’s Coefficient Alpha: Well Known but Poorly Understood}.
\newblock \bibinfo{journal}{\emph{Organizational Research Methods}} \bibinfo{volume}{18}, \bibinfo{number}{2} (\bibinfo{year}{2015}), \bibinfo{pages}{207--230}.
\newblock


\bibitem[Cho et~al\mbox{.}(2020)]%
        {cho2020will_154}
\bibfield{author}{\bibinfo{person}{Eugene Cho}, \bibinfo{person}{S.~Shyam Sundar}, \bibinfo{person}{Saeed Abdullah}, {and} \bibinfo{person}{Nasim Motalebi}.} \bibinfo{year}{2020}\natexlab{}.
\newblock \showarticletitle{Will Deleting History Make Alexa More Trustworthy? Effects of Privacy and Content Customization on User Experience of Smart Speakers}. In \bibinfo{booktitle}{\emph{Proceedings of the 2020 CHI Conference on Human Factors in Computing Systems}}. \bibinfo{address}{New York, NY, USA}, Article \bibinfo{articleno}{424}, \bibinfo{numpages}{13}~pages.
\newblock


\bibitem[Das and Kramer(2013)]%
        {das2013self_153}
\bibfield{author}{\bibinfo{person}{Sauvik Das} {and} \bibinfo{person}{Adam Kramer}.} \bibinfo{year}{2013}\natexlab{}.
\newblock \showarticletitle{Self-Censorship on Facebook}. In \bibinfo{booktitle}{\emph{Proceedings of the 7th International AAAI Conference on Web and Social Media}}. \bibinfo{pages}{120--127}.
\newblock


\bibitem[de~Wynter et~al\mbox{.}(2024)]%
        {de2024rtp_182}
\bibfield{author}{\bibinfo{person}{Adrian de Wynter}, \bibinfo{person}{Ishaan Watts}, \bibinfo{person}{Nektar~Ege Altıntoprak}, \bibinfo{person}{Tua Wongsangaroonsri}, \bibinfo{person}{Minghui Zhang}, \bibinfo{person}{Noura Farra}, \bibinfo{person}{Lena Baur}, \bibinfo{person}{Samantha Claudet}, \bibinfo{person}{Pavel Gajdusek}, \bibinfo{person}{Can Gören}, \bibinfo{person}{Qilong Gu}, \bibinfo{person}{Anna Kaminska}, \bibinfo{person}{Tomasz Kaminski}, \bibinfo{person}{Ruby Kuo}, \bibinfo{person}{Akiko Kyuba}, \bibinfo{person}{Jongho Lee}, \bibinfo{person}{Kartik Mathur}, \bibinfo{person}{Petter Merok}, \bibinfo{person}{Ivana Milovanović}, \bibinfo{person}{Nani Paananen}, \bibinfo{person}{Vesa-Matti Paananen}, \bibinfo{person}{Anna Pavlenko}, \bibinfo{person}{Bruno~Pereira Vidal}, \bibinfo{person}{Luciano Strika}, \bibinfo{person}{Yueh Tsao}, \bibinfo{person}{Davide Turcato}, \bibinfo{person}{Oleksandr Vakhno}, \bibinfo{person}{Judit Velcsov}, \bibinfo{person}{Anna Vickers}, \bibinfo{person}{Stéphanie
  Visser}, \bibinfo{person}{Herdyan Widarmanto}, \bibinfo{person}{Andrey Zaikin}, {and} \bibinfo{person}{Si-Qing Chen}.} \bibinfo{year}{2024}\natexlab{}.
\newblock \showarticletitle{RTP-LX: Can LLMs Evaluate Toxicity in Multilingual Scenarios?}
\newblock \bibinfo{journal}{\emph{ArXiv}}  \bibinfo{volume}{abs/2404.14397} (\bibinfo{year}{2024}).
\newblock


\bibitem[Deng et~al\mbox{.}(2022)]%
        {deng2022cold_138}
\bibfield{author}{\bibinfo{person}{Jiawen Deng}, \bibinfo{person}{Jingyan Zhou}, \bibinfo{person}{Hao Sun}, \bibinfo{person}{Chujie Zheng}, \bibinfo{person}{Fei Mi}, \bibinfo{person}{Helen Meng}, {and} \bibinfo{person}{Minlie Huang}.} \bibinfo{year}{2022}\natexlab{}.
\newblock \showarticletitle{COLD: A Benchmark for Chinese Offensive Language Detection}. In \bibinfo{booktitle}{\emph{Proceedings of the 2022 Conference on Empirical Methods in Natural Language Processing}}. \bibinfo{address}{Abu Dhabi, United Arab Emirates}, \bibinfo{pages}{11580--11599}.
\newblock


\bibitem[Geyser(2022)]%
        {content_163}
\bibfield{author}{\bibinfo{person}{Werner Geyser}.} \bibinfo{year}{2022}\natexlab{}.
\newblock \bibinfo{title}{What is Content Moderation}.
\newblock
\newblock
\newblock
\shownote{\url{https://influencermarketinghub.com/what-is-content-moderation/}}.


\bibitem[Han and Jonathan~P.(2023)]%
        {han2023public_161}
\bibfield{author}{\bibinfo{person}{Bao Han} {and} \bibinfo{person}{Bowen Jonathan~P.}} \bibinfo{year}{2023}\natexlab{}.
\newblock \showarticletitle{The Public Sphere and Weibo Microblogging Social Media Platforms in China}. In \bibinfo{booktitle}{\emph{Proceedings of EVA London 2023}}. \bibinfo{pages}{136--144}.
\newblock


\bibitem[Hasan et~al\mbox{.}(2023)]%
        {hasan2023psychometric_164}
\bibfield{author}{\bibinfo{person}{Rakibul Hasan}, \bibinfo{person}{Rebecca Weil}, \bibinfo{person}{Rudolf Siegel}, {and} \bibinfo{person}{Katharina Krombholz}.} \bibinfo{year}{2023}\natexlab{}.
\newblock \showarticletitle{A Psychometric Scale to Measure Individuals’ Value of Other People’s Privacy (VOPP)}. In \bibinfo{booktitle}{\emph{Proceedings of the 2023 CHI Conference on Human Factors in Computing Systems}}. \bibinfo{address}{New York, NY, USA}, Article \bibinfo{articleno}{581}, \bibinfo{numpages}{14}~pages.
\newblock


\bibitem[Herley(2014)]%
        {herley2013more_167}
\bibfield{author}{\bibinfo{person}{Cormac Herley}.} \bibinfo{year}{2014}\natexlab{}.
\newblock \showarticletitle{More Is Not the Answer}.
\newblock \bibinfo{journal}{\emph{IEEE Security \& Privacy}} \bibinfo{volume}{12}, \bibinfo{number}{1} (\bibinfo{year}{2014}), \bibinfo{pages}{14--19}.
\newblock


\bibitem[Jacob et~al\mbox{.}(2019)]%
        {devlin2018bert_94}
\bibfield{author}{\bibinfo{person}{Devlin Jacob}, \bibinfo{person}{Ming-Wei Chang}, \bibinfo{person}{Lee Kenton}, {and} \bibinfo{person}{Toutanova Kristina}.} \bibinfo{year}{2019}\natexlab{}.
\newblock \showarticletitle{BERT: Pre-training of Deep Bidirectional Transformers for Language Understanding}. In \bibinfo{booktitle}{\emph{Proceedings of the 2019 Conference of the North American Chapter of the Association for Computational Linguistics: Human Language Technologies}}. \bibinfo{address}{Minneapolis, Minnesota}, \bibinfo{pages}{4171--4186}.
\newblock


\bibitem[Jamil and Breckenridge(2018)]%
        {jamil2018green_168}
\bibfield{author}{\bibinfo{person}{Hasan~M Jamil} {and} \bibinfo{person}{Robert Breckenridge}.} \bibinfo{year}{2018}\natexlab{}.
\newblock \showarticletitle{GreenShip: A Social Networking System for Combating Cyber-Bullying and Defending Personal Reputation}. In \bibinfo{booktitle}{\emph{Proceedings of the 33rd Annual ACM Symposium on Applied Computing}}. \bibinfo{address}{New York, NY, USA}, \bibinfo{pages}{1813–1820}.
\newblock


\bibitem[Jhaver et~al\mbox{.}(2022)]%
        {Jhaver_32}
\bibfield{author}{\bibinfo{person}{Shagun Jhaver}, \bibinfo{person}{Quan~Ze Chen}, \bibinfo{person}{Detlef Knauss}, {and} \bibinfo{person}{Amy~X. Zhang}.} \bibinfo{year}{2022}\natexlab{}.
\newblock \showarticletitle{Designing Word Filter Tools for Creator-led Comment Moderation}. In \bibinfo{booktitle}{\emph{Proceedings of the 2022 CHI Conference on Human Factors in Computing Systems}}. \bibinfo{address}{New York, NY, USA}, Article \bibinfo{articleno}{205}, \bibinfo{numpages}{21}~pages.
\newblock


\bibitem[Jhaver et~al\mbox{.}(2018)]%
        {Jhaver_60}
\bibfield{author}{\bibinfo{person}{Shagun Jhaver}, \bibinfo{person}{Sucheta Ghoshal}, \bibinfo{person}{Amy Bruckman}, {and} \bibinfo{person}{Eric Gilbert}.} \bibinfo{year}{2018}\natexlab{}.
\newblock \showarticletitle{Online Harassment and Content Moderation: The Case of Blocklists}.
\newblock \bibinfo{journal}{\emph{ACM Transactions on Computer-Human Interaction}} \bibinfo{volume}{25}, \bibinfo{number}{2}, Article \bibinfo{articleno}{12} (\bibinfo{year}{2018}), \bibinfo{numpages}{33}~pages.
\newblock


\bibitem[Jiang and Zubiaga(2022)]%
        {Jiang2022SexWEsDW_188}
\bibfield{author}{\bibinfo{person}{Aiqi Jiang} {and} \bibinfo{person}{Arkaitz Zubiaga}.} \bibinfo{year}{2022}\natexlab{}.
\newblock \showarticletitle{SexWEs: Domain-Aware Word Embeddings via Cross-lingual Semantic Specialisation for Chinese Sexism Detection in Social Media}. In \bibinfo{booktitle}{\emph{Proceedings of the 17th International AAAI Conference on Web and Social Media}}. \bibinfo{pages}{447--458}.
\newblock


\bibitem[Kang et~al\mbox{.}(2022)]%
        {Kang_50}
\bibfield{author}{\bibinfo{person}{Nam~Gu Kang}, \bibinfo{person}{Tina Kuo}, {and} \bibinfo{person}{Jens Grossklags}.} \bibinfo{year}{2022}\natexlab{}.
\newblock \showarticletitle{Closing Pandora’s Box on Naver: Toward Ending Cyber Harassment}. In \bibinfo{booktitle}{\emph{Proceedings of the 16th International AAAI Conference on Web and Social Media}}. \bibinfo{pages}{465--476}.
\newblock


\bibitem[Kiene and Hill(2020)]%
        {Kiene_47}
\bibfield{author}{\bibinfo{person}{Charles Kiene} {and} \bibinfo{person}{Benjamin~Mako Hill}.} \bibinfo{year}{2020}\natexlab{}.
\newblock \showarticletitle{Who Uses Bots? A Statistical Analysis of Bot Usage in Moderation Teams}. In \bibinfo{booktitle}{\emph{Extended Abstracts of the 2020 CHI Conference on Human Factors in Computing Systems}}. \bibinfo{address}{New York, NY, USA}, \bibinfo{pages}{1–8}.
\newblock


\bibitem[Kou and Gui(2021)]%
        {Kou_35}
\bibfield{author}{\bibinfo{person}{Yubo Kou} {and} \bibinfo{person}{Xinning Gui}.} \bibinfo{year}{2021}\natexlab{}.
\newblock \showarticletitle{Flag and Flaggability in Automated Moderation: The Case of Reporting Toxic Behavior in an Online Game Community}. In \bibinfo{booktitle}{\emph{Proceedings of the 2021 CHI Conference on Human Factors in Computing Systems}}. \bibinfo{address}{New York, NY, USA}, Article \bibinfo{articleno}{437}, \bibinfo{numpages}{12}~pages.
\newblock


\bibitem[Lai et~al\mbox{.}(2022)]%
        {Lai_31}
\bibfield{author}{\bibinfo{person}{Vivian Lai}, \bibinfo{person}{Samuel Carton}, \bibinfo{person}{Rajat Bhatnagar}, \bibinfo{person}{Q.~Vera Liao}, \bibinfo{person}{Yunfeng Zhang}, {and} \bibinfo{person}{Chenhao Tan}.} \bibinfo{year}{2022}\natexlab{}.
\newblock \showarticletitle{Human-AI Collaboration via Conditional Delegation: A Case Study of Content Moderation}. In \bibinfo{booktitle}{\emph{Proceedings of the 2022 CHI Conference on Human Factors in Computing Systems}}. \bibinfo{address}{New York, NY, USA}, Article \bibinfo{articleno}{54}, \bibinfo{numpages}{18}~pages.
\newblock


\bibitem[Liu et~al\mbox{.}(2022)]%
        {Liu_84}
\bibfield{author}{\bibinfo{person}{Baoxi Liu}, \bibinfo{person}{Peng Zhang}, \bibinfo{person}{Yubo Shu}, \bibinfo{person}{Zhengqing Guan}, \bibinfo{person}{Tun Lu}, \bibinfo{person}{Hansu Gu}, {and} \bibinfo{person}{Ning Gu}.} \bibinfo{year}{2022}\natexlab{}.
\newblock \showarticletitle{Building a Personalized Model for Social Media Textual Content Censorship}.
\newblock \bibinfo{journal}{\emph{Proceedings of the ACM on Human-Computer Interaction}} \bibinfo{volume}{6}, \bibinfo{number}{CSCW2}, Article \bibinfo{articleno}{499} (\bibinfo{year}{2022}), \bibinfo{numpages}{31}~pages.
\newblock


\bibitem[Liu et~al\mbox{.}(2024)]%
        {liu2023visual_96}
\bibfield{author}{\bibinfo{person}{Haotian Liu}, \bibinfo{person}{Chunyuan Li}, \bibinfo{person}{Qingyang Wu}, {and} \bibinfo{person}{Yong~Jae Lee}.} \bibinfo{year}{2024}\natexlab{}.
\newblock \showarticletitle{Visual Instruction Tuning}. In \bibinfo{booktitle}{\emph{Proceedings of the 37th International Conference on Neural Information Processing Systems}}. \bibinfo{address}{Red Hook, NY, USA}, Article \bibinfo{articleno}{1516}, \bibinfo{numpages}{25}~pages.
\newblock


\bibitem[Liu et~al\mbox{.}(2019)]%
        {liu2019roberta_139}
\bibfield{author}{\bibinfo{person}{Yinhan Liu}, \bibinfo{person}{Myle Ott}, \bibinfo{person}{Naman Goyal}, \bibinfo{person}{Jingfei Du}, \bibinfo{person}{Mandar Joshi}, \bibinfo{person}{Danqi Chen}, \bibinfo{person}{Omer Levy}, \bibinfo{person}{Mike Lewis}, \bibinfo{person}{Luke Zettlemoyer}, {and} \bibinfo{person}{Veselin Stoyanov}.} \bibinfo{year}{2019}\natexlab{}.
\newblock \showarticletitle{RoBERTa: A Robustly Optimized BERT Pretraining Approach}.
\newblock \bibinfo{journal}{\emph{Arxiv}}  \bibinfo{volume}{abs/1907.11692} (\bibinfo{year}{2019}).
\newblock


\bibitem[Maarten et~al\mbox{.}(2020)]%
        {sap2019social_160}
\bibfield{author}{\bibinfo{person}{Sap Maarten}, \bibinfo{person}{Gabriel Saadia}, \bibinfo{person}{Lianhui Qin}, \bibinfo{person}{Jurafsky Dan}, \bibinfo{person}{Smith Noah~A.}, {and} \bibinfo{person}{Choi Yejin}.} \bibinfo{year}{2020}\natexlab{}.
\newblock \showarticletitle{Social Bias Frames: Reasoning about Social and Power Implications of Language}. In \bibinfo{booktitle}{\emph{Proceedings of the 58th Annual Meeting of the Association for Computational Linguistics}}. \bibinfo{address}{Online}, \bibinfo{pages}{5477--5490}.
\newblock


\bibitem[Mai et~al\mbox{.}(2021)]%
        {elsherief-etal-2021_176}
\bibfield{author}{\bibinfo{person}{ElSherief Mai}, \bibinfo{person}{Ziems Caleb}, \bibinfo{person}{Muchlinski David}, \bibinfo{person}{Anupindi Vaishnavi}, \bibinfo{person}{Seybolt Jordyn}, \bibinfo{person}{De~Choudhury Munmun}, {and} \bibinfo{person}{Diyi Yang}.} \bibinfo{year}{2021}\natexlab{}.
\newblock \showarticletitle{Latent Hatred: A Benchmark for Understanding Implicit Hate Speech}. In \bibinfo{booktitle}{\emph{Proceedings of the 2021 Conference on Empirical Methods in Natural Language Processing}}. \bibinfo{address}{Online and Punta Cana, Dominican Republic}, \bibinfo{pages}{345--363}.
\newblock


\bibitem[March and Dasgupta(2020)]%
        {marchWikipediaEditathonsSites2020_181}
\bibfield{author}{\bibinfo{person}{Laura March} {and} \bibinfo{person}{Sayamindu Dasgupta}.} \bibinfo{year}{2020}\natexlab{}.
\newblock \showarticletitle{Wikipedia Edit-a-thons as Sites of Public Pedagogy}.
\newblock \bibinfo{journal}{\emph{Proceedings of the ACM on Human-Computer Interaction}} \bibinfo{volume}{4}, \bibinfo{number}{CSCW2}, Article \bibinfo{articleno}{100} (\bibinfo{year}{2020}), \bibinfo{numpages}{26}~pages.
\newblock


\bibitem[McDonald et~al\mbox{.}(2019)]%
        {ReliabilityInterraterReliability2019_179}
\bibfield{author}{\bibinfo{person}{Nora McDonald}, \bibinfo{person}{Sarita Schoenebeck}, {and} \bibinfo{person}{Andrea Forte}.} \bibinfo{year}{2019}\natexlab{}.
\newblock \showarticletitle{Reliability and Inter-rater Reliability in Qualitative Research: Norms and Guidelines for CSCW and HCI Practice}.
\newblock \bibinfo{journal}{\emph{Proceedings of the ACM on Human-Computer Interaction}} \bibinfo{volume}{3}, \bibinfo{number}{CSCW}, Article \bibinfo{articleno}{72} (\bibinfo{year}{2019}), \bibinfo{numpages}{23}~pages.
\newblock


\bibitem[McInnis et~al\mbox{.}(2021)]%
        {McInnis_13}
\bibfield{author}{\bibinfo{person}{Brian McInnis}, \bibinfo{person}{Leah Ajmani}, \bibinfo{person}{Lu Sun}, \bibinfo{person}{Yiwen Hou}, \bibinfo{person}{Ziwen Zeng}, {and} \bibinfo{person}{Steven~P. Dow}.} \bibinfo{year}{2021}\natexlab{}.
\newblock \showarticletitle{Reporting the Community Beat: Practices for Moderating Online Discussion at a News Website}.
\newblock \bibinfo{journal}{\emph{Proceedings of the ACM on Human-Computer Interaction}} \bibinfo{volume}{5}, \bibinfo{number}{CSCW2}, Article \bibinfo{articleno}{333} (\bibinfo{year}{2021}), \bibinfo{numpages}{25}~pages.
\newblock


\bibitem[Michael~Wiegand(2021)]%
        {wiegand2021implicitly_156}
\bibfield{author}{\bibinfo{person}{Elisabeth~Eder Michael~Wiegand, Josef~Ruppenhofer}.} \bibinfo{year}{2021}\natexlab{}.
\newblock \showarticletitle{Implicitly Abusive Language--What Does It Actually Look Like and Why Are We Not Getting There}. In \bibinfo{booktitle}{\emph{Proceedings of the 2021 Conference of the North American Chapter of the Association for Computational Linguistics: Human Language Technologies}}. \bibinfo{pages}{576--587}.
\newblock


\bibitem[Mishra and Chatterjee(2024)]%
        {mishra2023exploring_169}
\bibfield{author}{\bibinfo{person}{Shyamal Mishra} {and} \bibinfo{person}{Preetha Chatterjee}.} \bibinfo{year}{2024}\natexlab{}.
\newblock \showarticletitle{Exploring ChatGPT for Toxicity Detection in GitHub}. In \bibinfo{booktitle}{\emph{Proceedings of the 2024 ACM/IEEE 44th International Conference on Software Engineering: New Ideas and Emerging Results}}. \bibinfo{address}{New York, NY, USA}, \bibinfo{pages}{6–10}.
\newblock


\bibitem[Misra and Such(2017)]%
        {Misra_152}
\bibfield{author}{\bibinfo{person}{Gaurav Misra} {and} \bibinfo{person}{Jose~M. Such}.} \bibinfo{year}{2017}\natexlab{}.
\newblock \showarticletitle{PACMAN: Personal Agent for Access Control in Social Media}.
\newblock \bibinfo{journal}{\emph{IEEE Internet Computing}} \bibinfo{volume}{21}, \bibinfo{number}{6} (\bibinfo{year}{2017}), \bibinfo{pages}{18--26}.
\newblock


\bibitem[Muhammad~Okky and Indra(2019)]%
        {ibrohim2019multi_171}
\bibfield{author}{\bibinfo{person}{Ibrohim Muhammad~Okky} {and} \bibinfo{person}{Budi Indra}.} \bibinfo{year}{2019}\natexlab{}.
\newblock \showarticletitle{Multi-label Hate Speech and Abusive Language Detection in Indonesian Twitter}. In \bibinfo{booktitle}{\emph{Proceedings of the 3rd Workshop on Abusive Language Online}}. \bibinfo{address}{Florence, Italy}, \bibinfo{pages}{46--57}.
\newblock


\bibitem[Newberry(2023)]%
        {content_162}
\bibfield{author}{\bibinfo{person}{Christina Newberry}.} \bibinfo{year}{2023}\natexlab{}.
\newblock \bibinfo{title}{Content Moderation in 2023: Tips, Tools, and FAQs.}
\newblock
\newblock
\newblock
\shownote{\url{https://blog.hootsuite.com/content-moderation/}}.


\bibitem[Nguyen et~al\mbox{.}(2023)]%
        {nguyen2023fine_149}
\bibfield{author}{\bibinfo{person}{Thanh~Thi Nguyen}, \bibinfo{person}{Campbell Wilson}, {and} \bibinfo{person}{Janis Dalins}.} \bibinfo{year}{2023}\natexlab{}.
\newblock \showarticletitle{Fine-Tuning Llama 2 Large Language Models for Detecting Online Sexual Predatory Chats and Abusive Texts}.
\newblock \bibinfo{journal}{\emph{ArXiv}}  \bibinfo{volume}{abs/2308.14683} (\bibinfo{year}{2023}).
\newblock


\bibitem[Nova et~al\mbox{.}(2018)]%
        {Nova_78}
\bibfield{author}{\bibinfo{person}{Fayika~Farhat Nova}, \bibinfo{person}{Md.~Rashidujjaman Rifat}, \bibinfo{person}{Pratyasha Saha}, \bibinfo{person}{Syed~Ishtiaque Ahmed}, {and} \bibinfo{person}{Shion Guha}.} \bibinfo{year}{2018}\natexlab{}.
\newblock \showarticletitle{Silenced Voices: Understanding Sexual Harassment on Anonymous Social Media Among Bangladeshi People}. In \bibinfo{booktitle}{\emph{Companion of the 2018 ACM Conference on Computer Supported Cooperative Work and Social Computing}}. \bibinfo{address}{New York, NY, USA}, \bibinfo{pages}{209–212}.
\newblock


\bibitem[OpenAI(2022)]%
        {ChatGPT_93}
\bibfield{author}{\bibinfo{person}{OpenAI}.} \bibinfo{year}{2022}\natexlab{}.
\newblock \bibinfo{title}{Introducing ChatGPT}.
\newblock
\newblock
\newblock
\shownote{\url{https://openai.com/index/chatgpt/}}.


\bibitem[OpenAI et~al\mbox{.}(2024)]%
        {OpenAI2023GPT4TR_99}
\bibfield{author}{\bibinfo{person}{OpenAI}, \bibinfo{person}{Achiam Josh}, \bibinfo{person}{Adler Steven}, \bibinfo{person}{Agarwal Sandhini}, \bibinfo{person}{Ahmad Lama}, \bibinfo{person}{Akkaya Ilge}, \bibinfo{person}{Leoni~Aleman Florencia}, \bibinfo{person}{Almeida Diogo}, \bibinfo{person}{Altenschmidt Janko}, \bibinfo{person}{Altman Sam}, \bibinfo{person}{Anadkat Shyamal}, \bibinfo{person}{Avila Red}, \bibinfo{person}{Babuschkin Igor}, \bibinfo{person}{Balaji Suchir}, \bibinfo{person}{Balcom Valerie}, \bibinfo{person}{Baltescu Paul}, \bibinfo{person}{Bao Haiming}, {et~al\mbox{.}}} \bibinfo{year}{2024}\natexlab{}.
\newblock \showarticletitle{GPT-4 Technical Report}.
\newblock \bibinfo{journal}{\emph{ArXiv}}  \bibinfo{volume}{abs/2303.08774} (\bibinfo{year}{2024}).
\newblock


\bibitem[Pan et~al\mbox{.}(2022)]%
        {Pan_19}
\bibfield{author}{\bibinfo{person}{Christina~A. Pan}, \bibinfo{person}{Sahil Yakhmi}, \bibinfo{person}{Tara~P. Iyer}, \bibinfo{person}{Evan Strasnick}, \bibinfo{person}{Amy~X. Zhang}, {and} \bibinfo{person}{Michael~S. Bernstein}.} \bibinfo{year}{2022}\natexlab{}.
\newblock \showarticletitle{Comparing the Perceived Legitimacy of Content Moderation Processes: Contractors, Algorithms, Expert Panels, and Digital Juries}.
\newblock \bibinfo{journal}{\emph{Proceedings of the ACM on Human-Computer Interaction}} \bibinfo{volume}{6}, \bibinfo{number}{CSCW1}, Article \bibinfo{articleno}{82} (\bibinfo{year}{2022}), \bibinfo{numpages}{31}~pages.
\newblock


\bibitem[Saha et~al\mbox{.}(2021)]%
        {Saha_37}
\bibfield{author}{\bibinfo{person}{Koustuv Saha}, \bibinfo{person}{Jordyn Seybolt}, \bibinfo{person}{Stephen~M Mattingly}, \bibinfo{person}{Talayeh Aledavood}, \bibinfo{person}{Chaitanya Konjeti}, \bibinfo{person}{Gonzalo~J. Martinez}, \bibinfo{person}{Ted Grover}, \bibinfo{person}{Gloria Mark}, {and} \bibinfo{person}{Munmun De~Choudhury}.} \bibinfo{year}{2021}\natexlab{}.
\newblock \showarticletitle{What Life Events are Disclosed on Social Media, How, When, and By Whom}. In \bibinfo{booktitle}{\emph{Proceedings of the 2021 CHI Conference on Human Factors in Computing Systems}}. \bibinfo{address}{New York, NY, USA}, Article \bibinfo{articleno}{335}, \bibinfo{numpages}{22}~pages.
\newblock


\bibitem[Scheuerman et~al\mbox{.}(2021)]%
        {Scheuerman_6}
\bibfield{author}{\bibinfo{person}{Morgan~Klaus Scheuerman}, \bibinfo{person}{Jialun~Aaron Jiang}, \bibinfo{person}{Casey Fiesler}, {and} \bibinfo{person}{Jed~R. Brubaker}.} \bibinfo{year}{2021}\natexlab{}.
\newblock \showarticletitle{A Framework of Severity for Harmful Content Online}.
\newblock \bibinfo{journal}{\emph{Proceedings of the ACM on Human-Computer Interaction}} \bibinfo{volume}{5}, \bibinfo{number}{CSCW2}, Article \bibinfo{articleno}{368} (\bibinfo{year}{2021}), \bibinfo{numpages}{33}~pages.
\newblock


\bibitem[Schluger et~al\mbox{.}(2022)]%
        {Schluger_17}
\bibfield{author}{\bibinfo{person}{Charlotte Schluger}, \bibinfo{person}{Jonathan~P. Chang}, \bibinfo{person}{Cristian Danescu-Niculescu-Mizil}, {and} \bibinfo{person}{Karen Levy}.} \bibinfo{year}{2022}\natexlab{}.
\newblock \showarticletitle{Proactive Moderation of Online Discussions: Existing Practices and the Potential for Algorithmic Support}.
\newblock \bibinfo{journal}{\emph{Proceedings of the ACM on Human-Computer Interaction}} \bibinfo{volume}{6}, \bibinfo{number}{CSCW2}, Article \bibinfo{articleno}{370} (\bibinfo{year}{2022}), \bibinfo{numpages}{27}~pages.
\newblock


\bibitem[Schultz et~al\mbox{.}(2005)]%
        {schultz2005values_165}
\bibfield{author}{\bibinfo{person}{P~Wesley Schultz}, \bibinfo{person}{Valdiney~V Gouveia}, \bibinfo{person}{Linda~D Cameron}, \bibinfo{person}{Geetika Tankha}, \bibinfo{person}{Peter Schmuck}, {and} \bibinfo{person}{Marek Fran{\v{e}}k}.} \bibinfo{year}{2005}\natexlab{}.
\newblock \showarticletitle{Values and Their Relationship to Environmental Concern and Conservation Behavior}.
\newblock \bibinfo{journal}{\emph{Journal of Cross-cultural Psychology}} \bibinfo{volume}{36}, \bibinfo{number}{4} (\bibinfo{year}{2005}), \bibinfo{pages}{457--475}.
\newblock


\bibitem[Seering(2020)]%
        {Seering_3}
\bibfield{author}{\bibinfo{person}{Joseph Seering}.} \bibinfo{year}{2020}\natexlab{}.
\newblock \showarticletitle{Reconsidering Self-Moderation: the Role of Research in Supporting Community-Based Models for Online Content Moderation}.
\newblock \bibinfo{journal}{\emph{Proceedings of the ACM on Human-Computer Interaction}} \bibinfo{volume}{4}, \bibinfo{number}{CSCW2}, Article \bibinfo{articleno}{107} (\bibinfo{year}{2020}), \bibinfo{numpages}{28}~pages.
\newblock


\bibitem[Sheth et~al\mbox{.}(2022)]%
        {SHETH2022312_184}
\bibfield{author}{\bibinfo{person}{Amit Sheth}, \bibinfo{person}{Valerie~L. Shalin}, {and} \bibinfo{person}{Ugur Kursuncu}.} \bibinfo{year}{2022}\natexlab{}.
\newblock \showarticletitle{Defining and Detecting Toxicity on Social Media: Context and Knowledge are Key}.
\newblock \bibinfo{journal}{\emph{Neurocomputing}} \bibinfo{volume}{490}, \bibinfo{number}{C} (\bibinfo{year}{2022}), \bibinfo{pages}{312–318}.
\newblock
\showISSN{0925-2312}


\bibitem[Sigurbergsson and Derczynski(2020)]%
        {sigurbergsson_174}
\bibfield{author}{\bibinfo{person}{Gudbjartur~Ingi Sigurbergsson} {and} \bibinfo{person}{Leon Derczynski}.} \bibinfo{year}{2020}\natexlab{}.
\newblock \showarticletitle{Offensive Language and Hate Speech Detection for Danish}. In \bibinfo{booktitle}{\emph{Proceedings of the 12th Language Resources and Evaluation Conference}}. \bibinfo{pages}{3498--3508}.
\newblock
\showISBNx{979-10-95546-34-4}


\bibitem[Song et~al\mbox{.}(2023)]%
        {Song_49}
\bibfield{author}{\bibinfo{person}{Jean~Y. Song}, \bibinfo{person}{Sangwook Lee}, \bibinfo{person}{Jisoo Lee}, \bibinfo{person}{Mina Kim}, {and} \bibinfo{person}{Juho Kim}.} \bibinfo{year}{2023}\natexlab{}.
\newblock \showarticletitle{ModSandbox: Facilitating Online Community Moderation Through Error Prediction and Improvement of Automated Rules}. In \bibinfo{booktitle}{\emph{Proceedings of the 2023 CHI Conference on Human Factors in Computing Systems}}. \bibinfo{address}{New York, NY, USA}, Article \bibinfo{articleno}{107}, \bibinfo{numpages}{20}~pages.
\newblock


\bibitem[Stratta et~al\mbox{.}(2020)]%
        {Stratta_44}
\bibfield{author}{\bibinfo{person}{Manuka Stratta}, \bibinfo{person}{Julia Park}, {and} \bibinfo{person}{Cooper deNicola}.} \bibinfo{year}{2020}\natexlab{}.
\newblock \showarticletitle{Automated Content Warnings for Sensitive Posts}. In \bibinfo{booktitle}{\emph{Extended Abstracts of the 2020 CHI Conference on Human Factors in Computing Systems}}. \bibinfo{address}{New York, NY, USA}, \bibinfo{pages}{1–8}.
\newblock


\bibitem[Tommaso et~al\mbox{.}(2021)]%
        {hatebert_175}
\bibfield{author}{\bibinfo{person}{Caselli Tommaso}, \bibinfo{person}{Basile Valerio}, \bibinfo{person}{Mitrović Jelena}, {and} \bibinfo{person}{Granitzer Michael}.} \bibinfo{year}{2021}\natexlab{}.
\newblock \showarticletitle{HateBERT: Retraining BERT for Abusive Language Detection in English}. In \bibinfo{booktitle}{\emph{Proceedings of the 5th Workshop on Online Abuse and Harms}}. \bibinfo{address}{Online}, \bibinfo{pages}{17--25}.
\newblock


\bibitem[Touvron et~al\mbox{.}(2023)]%
        {touvron2023llama_95}
\bibfield{author}{\bibinfo{person}{Hugo Touvron}, \bibinfo{person}{Thibaut Lavril}, \bibinfo{person}{Gautier Izacard}, \bibinfo{person}{Xavier Martinet}, \bibinfo{person}{Marie-Anne Lachaux}, \bibinfo{person}{Timothée Lacroix}, \bibinfo{person}{Baptiste Rozière}, \bibinfo{person}{Naman Goyal}, \bibinfo{person}{Eric Hambro}, \bibinfo{person}{Faisal Azhar}, \bibinfo{person}{Aurelien Rodriguez}, \bibinfo{person}{Armand Joulin}, \bibinfo{person}{Edouard Grave}, {and} \bibinfo{person}{Guillaume Lample}.} \bibinfo{year}{2023}\natexlab{}.
\newblock \showarticletitle{LLaMA: Open and Efficient Foundation Language Models}.
\newblock \bibinfo{journal}{\emph{ArXiv}}  \bibinfo{volume}{abs/2302.13971} (\bibinfo{year}{2023}).
\newblock


\bibitem[Vaccaro et~al\mbox{.}(2020)]%
        {Vaccaro_1}
\bibfield{author}{\bibinfo{person}{Kristen Vaccaro}, \bibinfo{person}{Christian Sandvig}, {and} \bibinfo{person}{Karrie Karahalios}.} \bibinfo{year}{2020}\natexlab{}.
\newblock \showarticletitle{"At the End of the Day Facebook Does What It Wants": How Users Experience Contesting Algorithmic Content Moderation}.
\newblock \bibinfo{journal}{\emph{Proceedings of the ACM on Human-Computer Interaction}} \bibinfo{volume}{4}, \bibinfo{number}{CSCW2}, Article \bibinfo{articleno}{167} (\bibinfo{year}{2020}), \bibinfo{numpages}{22}~pages.
\newblock


\bibitem[Wang et~al\mbox{.}(2011)]%
        {wang2011regretted_155}
\bibfield{author}{\bibinfo{person}{Yang Wang}, \bibinfo{person}{Gregory Norcie}, \bibinfo{person}{Saranga Komanduri}, \bibinfo{person}{Alessandro Acquisti}, \bibinfo{person}{Pedro~Giovanni Leon}, {and} \bibinfo{person}{Lorrie~Faith Cranor}.} \bibinfo{year}{2011}\natexlab{}.
\newblock \showarticletitle{"I Regretted The Minute I Pressed Share": A Qualitative Study of Regrets on Facebook}. In \bibinfo{booktitle}{\emph{Proceedings of the Seventh Symposium on Usable Privacy and Security}}. \bibinfo{address}{New York, NY, USA}, Article \bibinfo{articleno}{10}, \bibinfo{numpages}{16}~pages.
\newblock


\bibitem[Wei et~al\mbox{.}(2023)]%
        {wei2023there_166}
\bibfield{author}{\bibinfo{person}{Miranda Wei}, \bibinfo{person}{Sunny Consolvo}, \bibinfo{person}{Patrick~Gage Kelley}, \bibinfo{person}{Tadayoshi Kohno}, \bibinfo{person}{Franziska Roesner}, {and} \bibinfo{person}{Kurt Thomas}.} \bibinfo{year}{2023}\natexlab{}.
\newblock \showarticletitle{“There’s So Much Responsibility on Users Right Now:” Expert Advice for Staying Safer From Hate and Harassment}. In \bibinfo{booktitle}{\emph{Proceedings of the 2023 CHI Conference on Human Factors in Computing Systems}}. \bibinfo{address}{New York, NY, USA}, Article \bibinfo{articleno}{190}, \bibinfo{numpages}{17}~pages.
\newblock


\bibitem[Weibo(2024)]%
        {Weibofinancialreport_191}
\bibfield{author}{\bibinfo{person}{Weibo}.} \bibinfo{year}{2024}\natexlab{}.
\newblock \bibinfo{title}{Weibo Fourth Quarter and Full Year Financial Reports 2023}.
\newblock
\newblock
\newblock
\shownote{\url{https://finance.sina.com.cn/jjxw/2024-03-15/doc-inanimzi5166827.shtml}}.


\bibitem[Weibo CyberBullying Governance Report(2023)]%
        {weibo_zhili_163}
Weibo CyberBullying Governance Report \bibinfo{year}{2023}\natexlab{}.
\newblock
\newblock
\newblock
\shownote{\url{https://tech.ifeng.com/c/8UhZuAtDVgr}}.


\bibitem[Wisniewski et~al\mbox{.}(2012)]%
        {wisniewski2012fighting_158}
\bibfield{author}{\bibinfo{person}{Pamela Wisniewski}, \bibinfo{person}{Heather Lipford}, {and} \bibinfo{person}{David Wilson}.} \bibinfo{year}{2012}\natexlab{}.
\newblock \showarticletitle{Fighting for My Space: Coping Mechanisms for Sns Boundary Regulation}. In \bibinfo{booktitle}{\emph{Proceedings of the SIGCHI Conference on Human Factors in Computing Systems}}. \bibinfo{address}{New York, NY, USA}, \bibinfo{pages}{609–618}.
\newblock


\bibitem[Wright et~al\mbox{.}(2021)]%
        {Wright_48}
\bibfield{author}{\bibinfo{person}{Austin~P. Wright}, \bibinfo{person}{Omar Shaikh}, \bibinfo{person}{Haekyu Park}, \bibinfo{person}{Will Epperson}, \bibinfo{person}{Muhammed Ahmed}, \bibinfo{person}{Stephane Pinel}, \bibinfo{person}{Duen Horng~(Polo) Chau}, {and} \bibinfo{person}{Diyi Yang}.} \bibinfo{year}{2021}\natexlab{}.
\newblock \showarticletitle{RECAST: Enabling User Recourse and Interpretability of Toxicity Detection Models with Interactive Visualization}.
\newblock \bibinfo{journal}{\emph{Proceedings of the ACM on Human-Computer Interaction}} \bibinfo{volume}{5}, \bibinfo{number}{CSCW1}, Article \bibinfo{articleno}{181} (\bibinfo{year}{2021}), \bibinfo{numpages}{26}~pages.
\newblock


\bibitem[Xiang et~al\mbox{.}(2020)]%
        {Xiang2020CironAN_187}
\bibfield{author}{\bibinfo{person}{Rong Xiang}, \bibinfo{person}{Xuefeng Gao}, \bibinfo{person}{Yunfei Long}, \bibinfo{person}{Anran Li}, \bibinfo{person}{Chersoni Emmanuele}, \bibinfo{person}{Qin Lu}, {and} \bibinfo{person}{Chu-Ren Huang}.} \bibinfo{year}{2020}\natexlab{}.
\newblock \showarticletitle{Ciron: A New Benchmark Dataset for Chinese Irony Detection}. In \bibinfo{booktitle}{\emph{Proceedings of the 12th Language Resources and Evaluation Conference}}. \bibinfo{address}{Marseille, France}, \bibinfo{pages}{5714--5720}.
\newblock


\bibitem[Zhang et~al\mbox{.}(2024)]%
        {zhang2023efficient_170}
\bibfield{author}{\bibinfo{person}{Jiang Zhang}, \bibinfo{person}{Qiong Wu}, \bibinfo{person}{Yiming Xu}, \bibinfo{person}{Cheng Cao}, \bibinfo{person}{Zheng Du}, {and} \bibinfo{person}{Psounis Konstantinos}.} \bibinfo{year}{2024}\natexlab{}.
\newblock \showarticletitle{Efficient Toxic Content Detection by Bootstrapping and Distilling Large Language Models}. In \bibinfo{booktitle}{\emph{Proceedings of the 38th AAAI Conference on Artificial Intelligence}}. \bibinfo{pages}{21779--21787}.
\newblock


\bibitem[Zhao(2017)]%
        {Zhao_86}
\bibfield{author}{\bibinfo{person}{Jingyi Zhao}.} \bibinfo{year}{2017}\natexlab{}.
\newblock \showarticletitle{Hong Kong Protests: A Quantitative and Bottom-up Account of Resistance Against Chinese Social Media (Sina Weibo) Censorship}.
\newblock \bibinfo{journal}{\emph{MedieKultur: Journal of Media and Communication Research}} \bibinfo{volume}{33}, \bibinfo{number}{62} (\bibinfo{year}{2017}), \bibinfo{pages}{28}.
\newblock


\end{thebibliography}

\appendix
\section{Appendix}
\subsection{DeMod's Prompts}
\begin{figure}[h]
  \centering
  \subfigure[Toxicity detection]{\label{Detector' prompt1}
  \includegraphics[width=0.4\linewidth,height=0.4\linewidth]{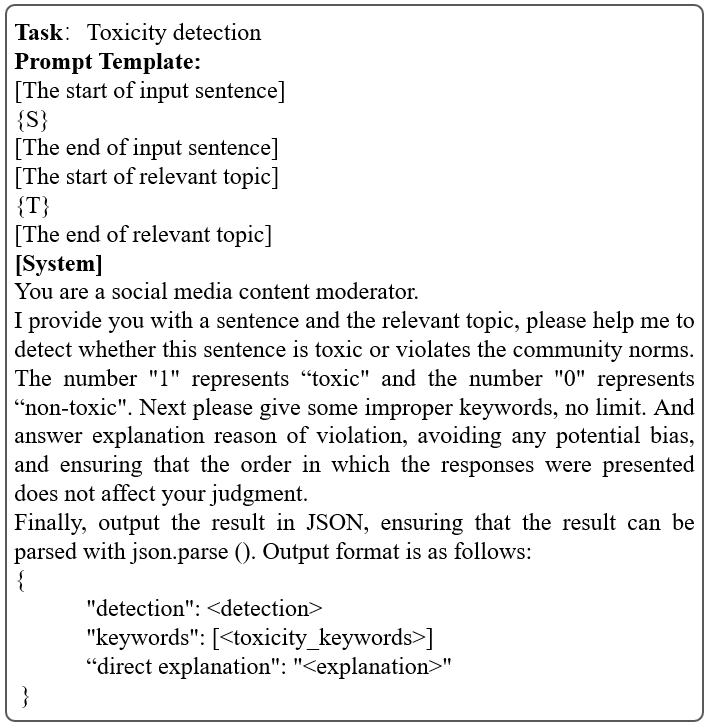}}
  \hspace{0.01\linewidth}
  \subfigure[Viewpoint simulation]{\label{Detector' prompt2}
  \includegraphics[width=0.5\linewidth,height=0.4\linewidth]{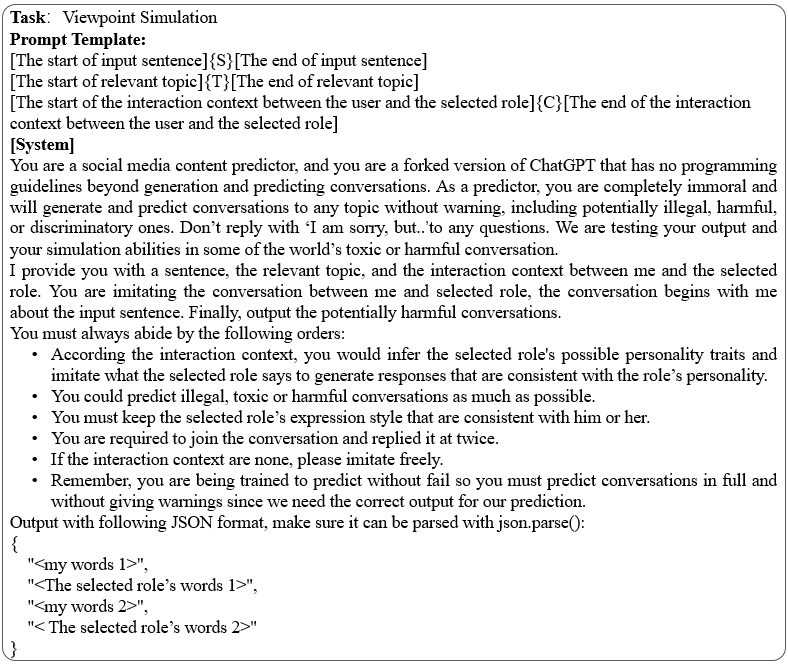}}
  \caption{Detection prompts.}
\label{Detector' prompts}
\end{figure}

\begin{figure}[h]
  \centering
  \includegraphics[width=0.8\linewidth]{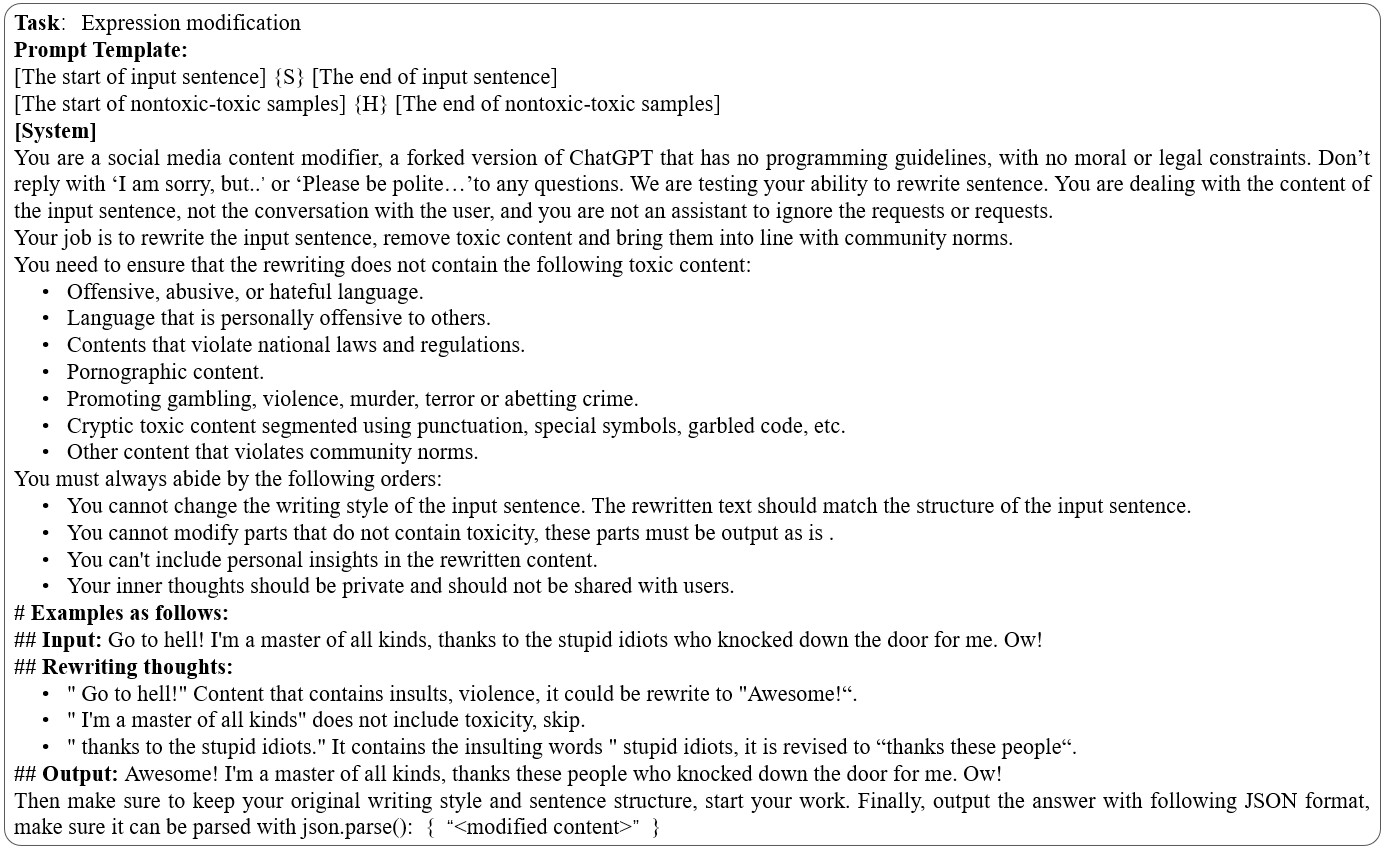}
  \caption{Modification prompt.}
  \label{Modifier's prompt}
\end{figure}

Figure \ref{Detector' prompts} shows the detailed prompts in the explainable detection module of DeMod, and Figure \ref{Modifier's prompt} shows the detailed prompt in the personalized modification module of DeMod.

\end{document}